\PassOptionsToPackage{dvipsnames}{xcolor}

\documentclass[sigplan,nonacm]{acmart}
\usepackage{amsmath}
\usepackage{enumitem}

\usepackage{array}
\usepackage{pifont}

\newcolumntype{L}[1]{>{\raggedright\arraybackslash}p{#1}}
\newcolumntype{B}[1]{>{\raggedright\arraybackslash}b{#1}}
\newcolumntype{R}[1]{>{\raggedleft\arraybackslash}p{#1}}

\usepackage[nounderscore]{syntax}

\usepackage{caption}
\usepackage{subcaption}

\usepackage{mathpartir}

\usepackage{multirow}

\usepackage{tikz}
\usetikzlibrary{positioning}

\definecolor{pasoBlue}{RGB}{46,134,193}
\definecolor{pasoGreen}{RGB}{39,174,96}
\definecolor{pasoRed}{RGB}{231,76,60}
\definecolor{pasoOrange}{RGB}{243,156,18}
\definecolor{pasoGray}{RGB}{127,140,141}
\definecolor{pasoPurple}{RGB}{142,68,173}
\definecolor{codeBg}{RGB}{245,246,250}
\definecolor{keywordBlue}{RGB}{26,82,118}

\colorlet{cb-blue}{RoyalBlue}
\colorlet{cb-green}{ForestGreen}
\colorlet{cb-pink}{CarnationPink}
\colorlet{cb-red}{RedOrange}
\colorlet{cb-grey}{gray}
\colorlet{cb-cyan}{Cyan}

\usepackage{listings, listings-rust}
\lstdefinelanguage{paso}{
  keywords={while, if, else, prot, fn, struct, in, out, step, fork,
  yield, assert\_eq, X, u1, u2, u8, u32, repeat, iterations, iters, uint},
  keywordstyle=\color{violet}\bfseries,
  otherkeywords={->,@},
  sensitive=true,
  comment=[l]{//}
}

\DeclareRobustCommand{\circled}[1]{%
  \tikz[baseline=(char.base)]{%
    \node[shape=circle, draw, inner sep=1pt, line width=0.4pt,
          font=\small\bfseries] (char) {#1};}}
\newcommand{\circleOne}{\circled{1}}
\newcommand{\circleTwo}{\circled{2}}
\newcommand{\circleThree}{\circled{3}}
\newcommand{\circleFour}{\circled{4}}

\definecolor{dkred}{rgb}{0.7,0,0}
\definecolor{dkgreen}{HTML}{006329}
\definecolor{dkpurple}{HTML}{4e02eb}

\newcount\draft\draft=1
\usepackage{xxx}

\def\code{\lstinline[basicstyle=\ttfamily]}

\def\lang{\mbox{\textsc{ProtoLang}}}

\newlength{\adderCodeH}
\newlength{\recvCodeH}
\newcommand{\monitorname}{reconstructor}

\author{Ernest Ng}
\affiliation{
  \institution{Cornell University}
  \city{Ithaca}
  \state{NY}
  \country{USA}
}
\email{eyn5@cornell.edu}

\author{Nikil V. Shyamsunder}
\affiliation{
  \institution{Cornell University}
  \city{Ithaca}
  \state{NY}
  \country{USA}
}
\email{nvs26@cornell.edu}

\author{Francis Pham}
\affiliation{
  \institution{Cornell University}
  \city{Ithaca}
  \state{NY}
  \country{USA}
}
\email{fdp25@cornell.edu}

\author{Adrian Sampson}
\affiliation{
  \institution{Cornell University}
  \city{Ithaca}
  \state{NY}
  \country{USA}
}
\email{asampson@cs.cornell.edu}

\author{Kevin Laeufer}
\affiliation{
  \institution{Cornell University}
  \city{Ithaca}
  \state{NY}
  \country{USA}
}
\email{laeufer@cornell.edu}

\title{Specifying Hardware Communication as Programs}

\begin{abstract}

To test and debug hardware modules, it is common to write two programs. A \emph{driver} maps high-level transactions to cycle-level I/O behavior, and a \emph{monitor} looks at the low-level I/O and recovers transactions. Both programs bridge the same abstraction gap between RTL and functional models.
Writing them separately
risks inconsistency.

\lang{} is a DSL for specifying hardware protocols.
The same \lang{} program can both drive and monitor transactions.
Users specify RTL communication protocols as succinct imperative programs. We present a novel \emph{\monitorname{}} algorithm that, given a \lang{} specification and a waveform, automatically infers a transaction-level trace consistent with the waveform. We demonstrate \lang{}'s utility via case studies focusing on the Wishbone and AXI-Stream interconnects, showing that \lang{} can model complex protocol features such as burst mode. We highlight how our tools can identify protocol violation bugs in real-world implementations.
\end{abstract}

\begin{document}
\maketitle

\makeatletter
\fancyhead[LO]{\ACM@linecountL}%
\fancyhead[RE]{\ACM@linecountR}%
\makeatother

\section{Introduction}
\label{sec:intro}

Hardware design at the register-transfer level (RTL) requires thinking at two levels of abstraction.
On one hand, every module has a set of high-level operations that it is supposed to implement.
An ALU should support addition, subtraction, and some bitwise operations;
a memory should support read and write transactions.
These \emph{architectural} specifications describe the set of \emph{transactions} that the module supports.
From another perspective, modules must carry out these transactions at the level of clock cycles, mapping every transaction onto a set of hardware signals over time.

This duality creates a gap between high-level functionality and low-level cycle-by-cycle behavior.
Methodologies for RTL design, integration, verification, testing, and debugging must bridge this gap.
We need to know \emph{when} and \emph{where} to provide a transaction's inputs and to expect its outputs.

A hardware \textit{protocol} is any strategy for mapping high-level transactions onto low-level RTL input/output behavior.
The spectrum includes simple handshakes, bespoke interfaces for internal modules,
and standardized buses like Wishbone and AMBA~\cite{Wishbone:2010,ARM:AMBA:1999}.
Protocols are often defined using informal prose.
Developers interpret these specifications when they write a unit test, interpret a waveform for debugging, or integrate with another RTL module.
We need a mechanism that lets us specify the protocol \emph{once} and then reuse it across many steps in the RTL design lifecycle.

Recent work has focused on type systems to encode protocols.
These type systems ensure that when an RTL module is used, the surrounding RTL modules will communicate with it according to its protocol.
However, type-based approaches generally require RTL to be rewritten in a new language.
Furthermore, they can only express certain kinds of protocols, such as statically timed pipelines~\cite{filament,lilac} or a pre-defined family of decoupled interfaces~\cite{yu2026anvil}.

Our goal is specify and check any existing hardware protocol without RTL code changes.
As a trade-off, we forgo any static guarantees and instead focus on dynamic testing and debugging.
We propose a domain-specific language (DSL), \lang{}, for describing protocols.
Without \lang{}, testing workflows encode protocol knowledge in two different pieces of code. The protocol \textit{driver} takes in the arguments to a high-level transaction and executes that transaction on the ports of the design under test (DUT). The protocol \textit{monitor} observes the ports of the DUT and recovers any transactions that it observes~\cite{uvm_manual}.
With \lang{}, developers specify a protocol once and use it for both driving and monitoring.

\lang{} borrows concepts from imperative testbench frameworks~\cite{cocotb, chiseltest,dobis2023verification,uvm_manual}.
The user writes a protocol definition as a function that takes the arguments of a transaction, such as the address and data for a \textit{write} transaction,
and manipulates the DUT's input and output ports.
We constrain \lang{} to make it feasible to use the same code as a monitor.
In other words, a \lang{} program runs in both ``directions'':
running it forward simulates transactions to produce a signal trace;
running it backward analyzes a signal trace to identify the transactions that explain the trace.

\begin{figure*}[t]
  \centering
  \includegraphics[width=\textwidth]{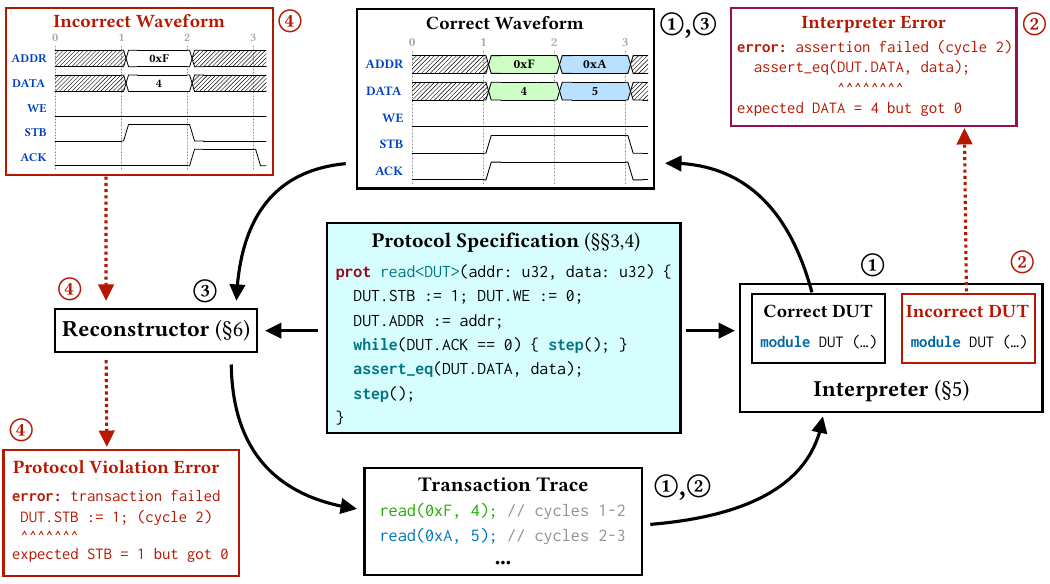}
  \caption{
    The \lang{} system, with a
    protocol describing a simplified Wishbone \code|read|.
    \circleOne\hspace{0.7pt} Given a \lang{} protocol, DUT, and sequence of transactions,
    our interpreter ($\S\ref{sec:the_paso_language}$) simulates the hardware for testing.
    \circleTwo\hspace{0.7pt} If the DUT is incorrect,
    the interpreter emits an error.
    \circleThree\hspace{0.7pt} Given a \lang{} protocol and a waveform, the \monitorname{} ($\S\ref{sec:monitor}$)
    infers a \emph{transaction trace} consistent with the waveform. This transaction trace indicates the cycles 
    when \code|read| transactions occurred.
    \circleFour\hspace{0.7pt} If the waveform is incorrect, i.e., it does not correspond to the protocol specification,
    the \monitorname{} emits an error. In the incorrect waveform shown, the \texttt{STB} signal is lowered 
    before \texttt{ACK} becomes high, which violates the classic Wishbone specification~\cite{Wishbone:2010}.
  }
  \label{fig:overview}
\end{figure*}

This paper's key contributions are:
\begin{itemize}
  \item We design a new DSL, \lang{}, for specifying hardware protocols.
  A single \lang{} program can both \emph{drive} hardware modules for testing and \emph{monitor} hardware executions for trace analysis ($\S\ref{sec:language_overview}$).
  \item We design an interpreter for \lang{} that avoids common pitfalls in \emph{ad hoc} Verilog testbenches, such as race conditions and erroneous sampling of transient values ($\S\ref{sec:the_paso_language}$).
  \item We present the \lang{} \emph{reconstructor}, a tool for automatically inferring transactions from a signal trace, even in the presence of control-flow and concurrency in \lang{} protocols ($\S\ref{sec:monitor}$).
  \item We demonstrate \lang{}'s utility via two case studies modeling complex features from the Wishbone interconnect and identifying protocol violation bugs in implementations of the AXI-Stream protocol ($\S\ref{sec:case_studies}$).
\end{itemize}

\section{Background}
\label{sec:background}

\paragraph{Synchronous RTL design}
This paper considers
digital, synchronous, register-transfer level (RTL) hardware modules.
In digital circuits,
every wire always contains a digital value (0 or 1) at the end of a clock cycle.
Under the RTL paradigm, 
designers explicitly define each state element (e.g., registers).
Synchrony means that state elements are always 
updated via a global clock event. 
\emph{Combinational} (non-state) elements do not incur any delay:
when their inputs change, their outputs are updated 
within the same cycle.
This paper measures all timing in cycles.
We do not consider designs with combinational loops (feedback loops with no state elements), which can cause unpredictable behavior~ \cite{altera_quartus_manual_2018}.

\paragraph{Testbenches, drivers and monitors}
To test an RTL module, one must drive values onto 
the module's input ports and check whether the value of output ports
are as expected. Moreover, one also needs to initialize 
state elements such as registers and elements. \emph{Testbenches}
are programs that interact with the DUT, allowing one 
to drive inputs and observe the DUT's outputs under simulation. 

Testbenches can be implemented using an HDL or using
libraries in software programming languages, such as Cocotb in Python \cite{cocotb}
and ChiselTest in Scala \cite{chiseltest}.
The Universal Verification
Methodology (UVM) \cite{uvm_manual} is a standard set of SystemVerilog
APIs for writing drivers and monitors.
However, no open-source hardware simulator supports
sufficient SV features for running UVM testbenches
\cite{laeufer_thesis_2024}.

\paragraph{Debugging signal traces}
\citet{ma_et_al_asplos_2022} found that within the 50 most 
popular FPGA projects on GitHub, 88\% do not include test cases 
in the form of executable testbench code. Instead, 
engineers most likely test their RTL designs in an ad-hoc manner
by visualizing \emph{signal traces} that are produced during simulation.
A signal trace records the value of DUT signals at every clock cycle, 
allowing engineers to inspect the cycle-level behavior of their designs.
These traces are visualized as \emph{waveforms} (also known as timing diagrams) 
using \textit{waveform viewers} (e.g.,
GTKWave \cite{gtkwave} or Surfer \cite{surfer_cav_2025}).

However, waveform debugging 
is a tedious and highly manual process
\cite{wal_aspdac2022, cider_asplos_2023, zhang2022_thesis}.
Recent studies have found that FPGA developers spend 
more time debugging than any other activity \cite{siemens_2020_fpga_report},
and a majority of them indicate a desire for better FPGA debugging tools \cite{exostiv_2015_fpga_report}.

\begin{figure*}[t] 
  \centering
  \begin{minipage}[t]{0.48\textwidth}
    \vspace{0pt}
    \centering
\begin{lstlisting}[language=paso,escapechar=!,aboveskip=0pt,belowskip=0pt]
prot add<DUT: Adder>(a: u32, b: u32, s: u32) {
  DUT.a := a; DUT.b := b; // Apply inputs
\end{lstlisting}%
\begin{lstlisting}[language=paso,aboveskip=0pt,belowskip=0pt,
  basicstyle=\ttfamily\fontsize{8.5}{10.2}\selectfont\color{pasoRed},
  keywordstyle=\color{pasoRed}\bfseries,
  commentstyle=\color{pasoRed}\ttfamily]
  // Pipelined variant only
  step();                 // One cycle delay
  DUT.a := X; DUT.b := X; // Release inputs
  fork();                 // Start next transaction
\end{lstlisting}%
\begin{lstlisting}[language=paso,aboveskip=0pt,belowskip=0pt]
  assert_eq(DUT.s, s);    // Check output
  step();                 // Finish cycle
}
\end{lstlisting}
  \end{minipage}%
  \hfill%
  \begin{minipage}[t]{0.48\textwidth}
    \vspace{0pt}
    \centering
    \begin{subfigure}[t]{0.48\linewidth}
      \centering
      \includegraphics[width=\linewidth]{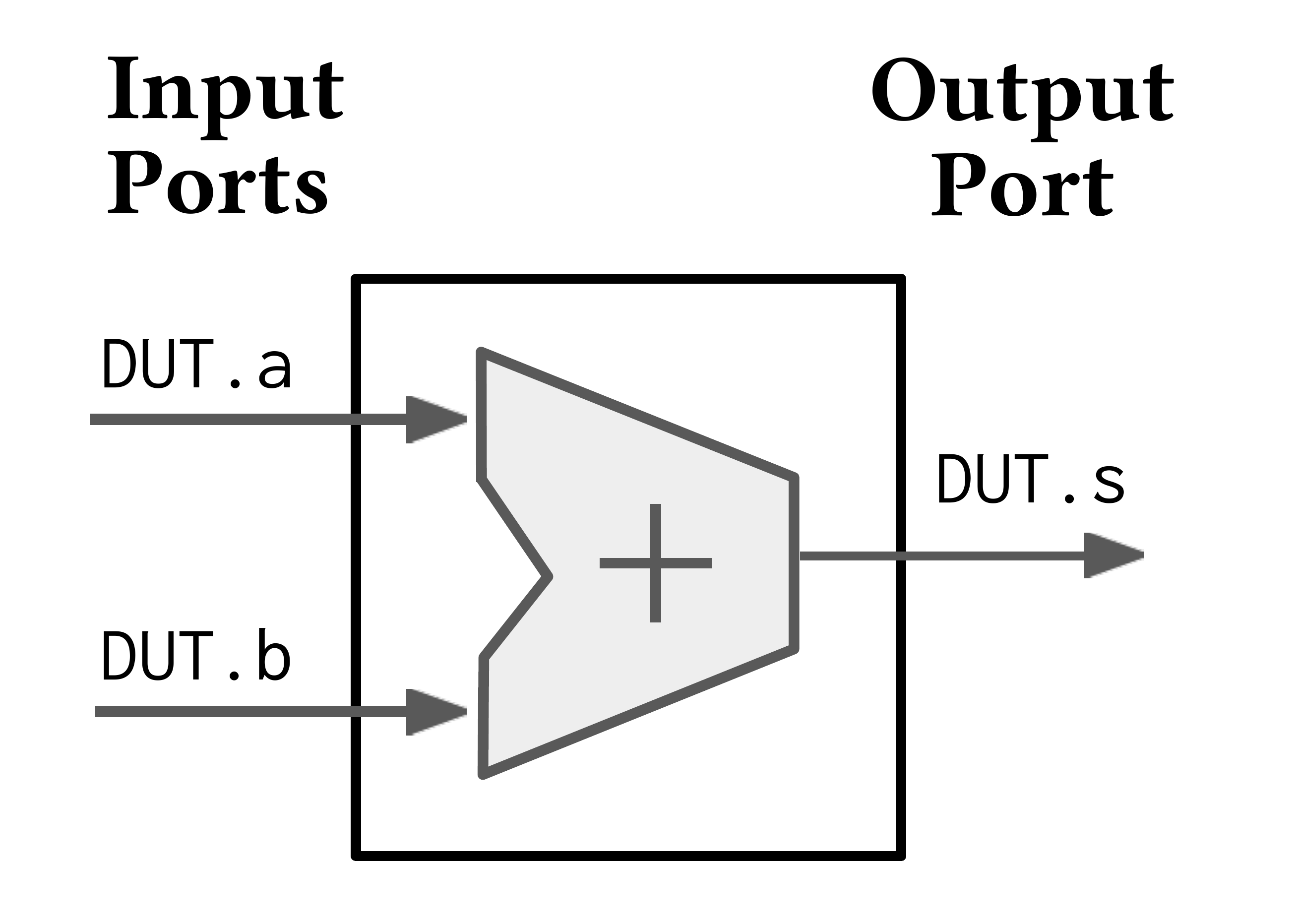}
      \caption{A combinational adder.}
      \label{fig:combinational_adder}
    \end{subfigure}
    \hfill
    \begin{subfigure}[t]{0.48\linewidth}
      \centering
      \includegraphics[width=\linewidth]{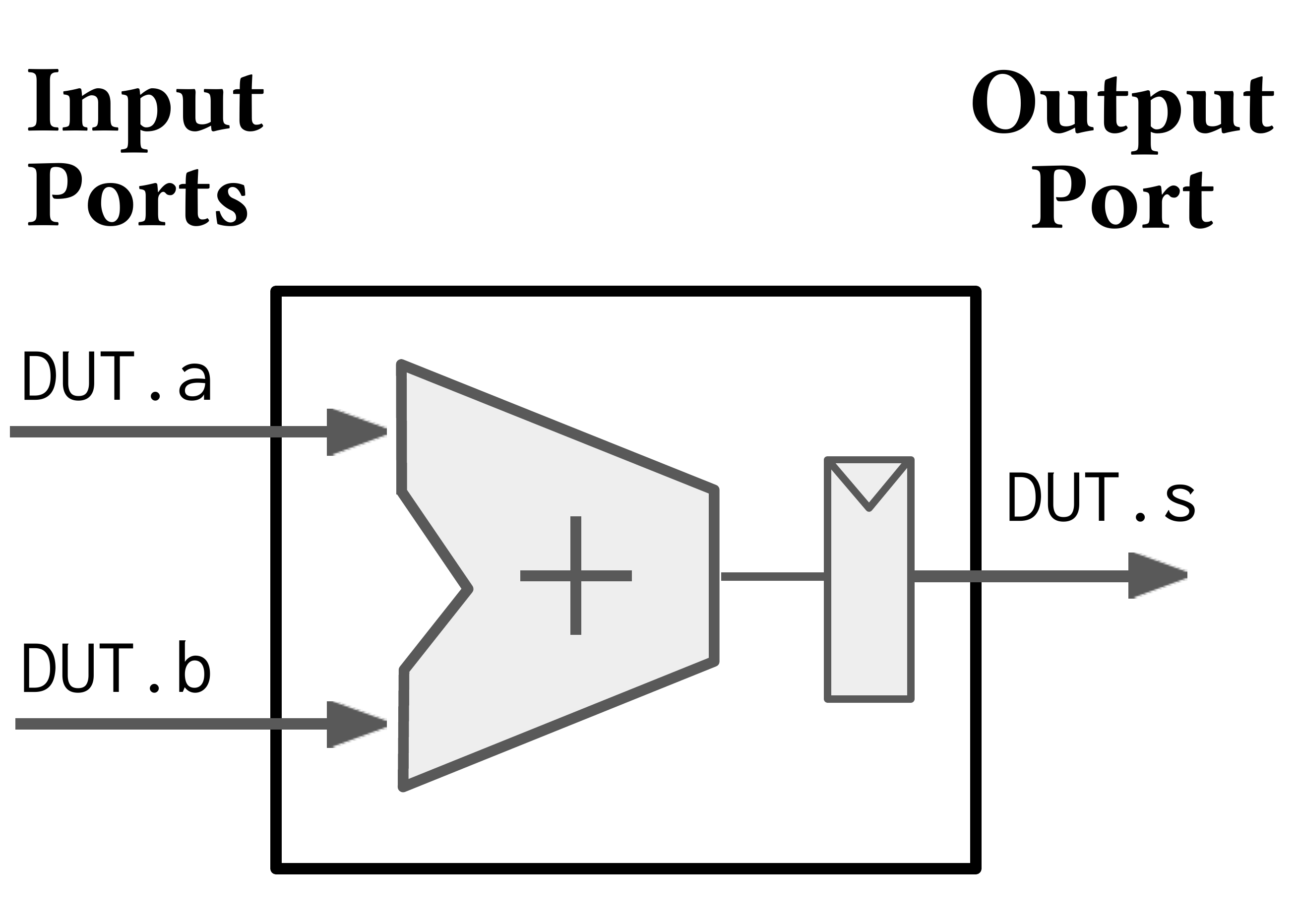}
      \caption{A pipelined adder.}
      \label{fig:pipelined_adder}
    \end{subfigure}
  \end{minipage}
  \caption{
    \label{fig:adder_protocols}
    A shared \lang{} adder protocol. The combinational implementation in
    Figure~\ref{fig:combinational_adder} uses only the un\textcolor{pasoRed}{highlighted} lines;
    the pipelined implementation in Figure~\ref{fig:pipelined_adder} also
    executes the \textcolor{pasoRed}{highlighted} lines.
  }
\end{figure*}

\section{Overview by Example}
\label{sec:language_overview}

This section introduces \lang{} via a series of examples.
\lang{} programs consist of \emph{protocols},
succinct imperative programs that describe the I/O behavior of hardware modules.
A \emph{transaction} is a dynamic instance of a protocol.

\lang{} protocols support two execution modes:
an \emph{interpreter} runs protocols ``forward'' to drive an RTL simulation,
and a \emph{reconstructor} runs them ``backward'' against a simulation trace to recover the transactions it reflects.
We describe sample use cases for each protocol to illustrate how these tools can aid debugging.

\subsection{Combinational Adder}

The un\textcolor{pasoRed}{highlighted} lines in Figure~\ref{fig:adder_protocols} show a \lang{} protocol that describes the interface to a simple combinational adder (Figure~\ref{fig:combinational_adder}).
The protocol declares \code|DUT| as a hardware module conforming to the \code|Adder| interface (not shown),
which specifies the required ports.
The arguments \texttt{a}, \texttt{b}, and \texttt{s} represent the data---both inputs and outputs---exchanged in each transaction.

The body of the protocol is an imperative program.
Assignments like \code|DUT.a := a| drive the design's input ports,
and \code|assert_eq| checks that outputs match an expected value.
(In the \monitorname{}, assignments and assertions give rise to constraints that it discharges when inferring transactions; 
we refer the reader to $\S\ref{sec:monitor}$ for further details.) 
Critically, protocols specify only a module's communication pattern, and not its functional behavior:
this protocol never asserts that \code|s| is the sum of \code|a| and \code|b|.

The \code|step| construct advances to the next clock cycle.
Every protocol in \lang{} must span at least one cycle, so it must contain at least one \code|step|.
\lang{} also prohibits assignments after the last \code|step| in a protocol to avoid conflicts between different transactions.

\paragraph{Use case: Transaction-level testing}
Consider simple concrete testing:
we want to write a series of test cases that check the adder's correctness on a few specific numbers.
One technique would be to write a Verilog or C++ testbench that applies pairs of operands and
checks the value of \code|s|.
However, this traditional approach conflates the functional specification (addition is numerically correct) with the communication protocol (when the input ports are driven, the sum is ready in the same cycle).
If we change the interface to our adder, as we will in the next section, we would need to change every test case to match.

Using \lang{}'s interpreter, however, developers can write \emph{transaction-level} testbenches.
To check the functional results for a series of transactions, we can write a transaction trace, e.g., \texttt{add(0, 42, 42); add(8, 12, 20)}.
Using this transaction trace, the interpreter executes two instantiations of the \code|add| protocol with the specified argument values.
This use case separates the functional specification (the transaction trace) from the communication details (the \lang{} program).

\subsection{Pipelined Adder}
\label{sec:pipelined_adder}

Figure~\ref{fig:pipelined_adder} shows a two-stage pipelined adder.
Figure~\ref{fig:adder_protocols}, including the \textcolor{pasoRed}{highlighted} lines, contains a modified \lang{} protocol that describes this interface, where outputs arrive in the next cycle after supplying the inputs.
The second \code|step| in the protocol advances the clock after the initial assignments to the input.
Because \code|assert_eq(DUT.s, s)| appears after the \code|step|, the protocol checks the output on the second cycle.

The second cycle also uses \lang{}'s \texttt{DontCare} assignments, written \code|DUT.a := X|.
These assignments reflect that, to use this adder, a client only needs to supply inputs in the first cycle, i.e., that the values of the DUT's \code|a| and \code|b| ports are irrelevant to the protocol at this point.

Finally, the protocol uses \lang{}'s \code|fork| primitive in the second cycle.
This construct indicates that another transaction may start concurrently at that point in the program.
Without \code|fork|, only a single transaction may be active at a time.
So the \code|fork| in this protocol enables pipelining:
one \code|add| transaction can start on each cycle.

\paragraph{Use case: Waveform analysis}

Imagine we discover a bug in a larger hardware design that includes this pipelined adder,
and we want to know whether the adder is to blame.
We might collect a signal trace that exhibits the bug and inspect the adder's input and output signals in a waveform viewer.
This could work, but it can be laborious to associate inputs with outputs.
For a pipeline depth of $P$, the output appears in the waveform $P$ cycles later than the inputs.

The \lang{} reconstructor can help associate the inputs with their corresponding outputs in the trace.
Using the protocol in Figure~\ref{fig:adder_protocols}, we can reconstruct every \code|add| transaction that occurs in the trace along with its \code|a|, \code|b|, and \code|s| arguments.
This transaction-level information can make it easier to identify instances where the inputs and outputs disagree.

\subsection{A Ready--Valid Handshake}
\label{sec:ready_valid_handshake}

The adder's protocol is simple because it is \emph{latency-sensitive} (LS), i.e., each transaction uses a specific, fixed number of clock cycles.
\emph{Latency-insensitive} (LI) interfaces are more complex because they use explicit signalling to allow dynamic latencies~\cite{carloni_et_al_2001}.

One example is a \emph{ready--valid} handshake,
which synchronizes a sender and a receiver.
A \code|data| port contains the payload.
The sender uses a \code|valid| bit to indicate when \code|data| contains a meaningful value;
the receiver uses a \code|ready| bit to indicate that it is ready to consume  new value.
When \code|valid| and \code|ready| are high simultaneously, a data transfer occurs.

\begin{figure}[t]
  \centering
\begin{lstlisting}[language=paso,escapechar=|,aboveskip=0pt,belowskip=0pt,
  numbers=left,numberstyle=\footnotesize\color{pasoGray},xleftmargin=2em]
prot recv<DUT: ReadyValid>(data: u8) {
  DUT.ready := 1; // Receiver asserts ready
  // Wait for valid data
  while (DUT.valid != 1) { step(); }
  assert_eq(D.data, data); // Data exchanged
  step(); // One cycle for data transfer
}
\end{lstlisting}%

    \vspace{0.75ex}

    \begin{subfigure}[b]{0.82\linewidth}
      \centering
      \includegraphics[width=\linewidth]{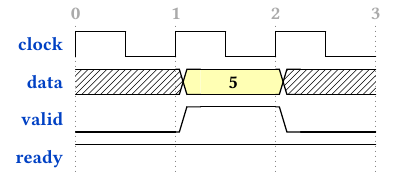}
      \caption{Receiver always drives ready.}
      \label{fig:ready_valid_protocol_always_ready}
    \end{subfigure}

  \vspace{1.5ex}

\begin{lstlisting}[language=paso,escapechar=|,aboveskip=0pt,belowskip=0pt,
  numbers=left,numberstyle=\footnotesize\color{pasoGray},xleftmargin=2em]
prot recv<DUT: ReadyValid>(data: u8, |\textcolor{pasoRed}{n: \textbf{uint}}|) {
  // Wait for valid data
  while (DUT.valid != 1) { step(); }
  assert_eq(D.data, data); // Data exchanged
\end{lstlisting}%
\begin{lstlisting}[language=paso,escapechar=|,aboveskip=0pt,belowskip=0pt,
  numbers=left,firstnumber=6,numberstyle=\footnotesize\color{pasoGray},xleftmargin=2em,
  basicstyle=\ttfamily\fontsize{8.5}{10.2}\selectfont\color{pasoRed},
  keywordstyle=\color{pasoRed}\bfseries,
  commentstyle=\color{pasoRed}\ttfamily]
  // Backpressure variant only
  DUT.ready := 0; // Apply backpressure
  repeat n iterations { 
    step();
    // While ready is low, valid must remain 
    // high and data may not change value
    assert_eq(DUT.valid, 1);
    assert_eq(DUT.data, data);
  }
\end{lstlisting}%
\begin{lstlisting}[language=paso,escapechar=|,aboveskip=0pt,belowskip=0pt,
  numbers=left,firstnumber=13,numberstyle=\footnotesize\color{pasoGray},xleftmargin=2em]
  DUT.ready := 1;
  step();
}
\end{lstlisting}

    \vspace{0.75ex}

    \begin{subfigure}[b]{0.82\linewidth}
      \centering
      \includegraphics[width=\linewidth]{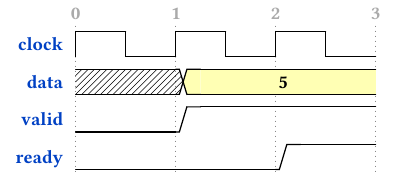}
      \caption{Receiver exerts backpressure for $n=1$ cycles.}
      \label{fig:ready_valid_protocol_with_backpressure}
    \end{subfigure}
  \caption{
    \label{fig:ready_valid_waveforms}
    Ready-valid receiver behaviors and their corresponding protocol code.
    The un\textcolor{pasoRed}{highlighted} statements produce waveform (a). Waveform (b)
    uses both un\textcolor{pasoRed}{highlighted} and \textcolor{pasoRed}{highlighted} backpressure statements for $n=1$.
  }
\end{figure}

We first consider a special case where receiver is always ready.
the un\textcolor{pasoRed}{highlighted} lines in Figure~\ref{fig:ready_valid_waveforms} show a \lang{} program describing this protocol,
and the waveform in Figure~\ref{fig:ready_valid_protocol_always_ready} illustrates one such transaction.
At the start of
this behavior, the
receiver immediately sets \texttt{DUT.ready := 1}, indicating that
it is always ready to receive data.
The \code|while| loop then waits for the \code|valid| signal to become high.
When it does, the \code|assert_eq| reflects the payload exchanged in that cycle.

\subsection{Ready--Valid With Backpressure}

In more sophisticated producer--consumer relationships, the receiver may not always be ready for new inputs.
By lowering the \code|ready| bit, a consumer can exert \emph{backpressure} to stall the producer.
The waveform in Figure~\ref{fig:ready_valid_protocol_with_backpressure} illustrates this situation: although the data is valid on cycle 1, a data transfer only occurs on cycle 2, when the receiver becomes ready.

The full code in Figure~\ref{fig:ready_valid_waveforms}, including the \textcolor{pasoRed}{highlighted} lines, generalizes the protocol to support backpressure.
The argument \code|n| determines the number of cycles that the transaction waits for the receiver to become ready.
To model this delay, the protocol sets \texttt{DUT.ready := 0},
then loops for \code|n| cycles. The \code|repeat| loop construct
is more restrictive than \code|while| loops (in general, 
\code|while|-loops cannot be emulated solely using \code|repeat| loops),
however it allows the number of backpressure cycles to be monitorable by our \monitorname{} ($\S\ref{sec:monitor}$).
Each iteration of the \code|repeat n| loop advances one clock cycle using \code|step|.
Ready--valid interfaces require data, when it becomes valid, to stay valid and constant until the transaction completes~\cite{arm_axi_stream_2021};
the \code|assert_eq| statements enforce this invariant.
After waiting \code|n| cycles, the protocol sets the \code|ready| flag and the transaction is complete. 
Here, one should consider \code|n| to be like an output of the transaction, as it describes the 
number of backpressure cycles that occur.

\paragraph{Use case: Detecting protocol violations}

Incorrect implementations of the ready--valid handshake can cause unpredictable behavior.
Using the \lang{} reconstructor, we can analyze a signal trace to check whether the protocol is running correctly.

The reconstructor identifies a sequence of transactions \code|recv(data, n)| that match a given signal trace.
For example, if a certain handshake in a trace waits for 9 cycles while $\texttt{valid} = 1$ and $\texttt{ready} = 0$, the reconstructor determines that the \code|repeat| loop in the protocol code must execute 9 times to explain the observed signals.
Therefore, in this transaction, $\texttt{n} = 9$.

Imagine, however, that the DUT's \code|data| signal changes its value on cycle 5.
The assertion \code|assert_eq(DUT.data, data)| requires the signal to be constant throughout the transaction.
Because there is no value of the parameter \code|data| that can make this assertion hold, the reconstructor determines that no valid \code|recv| transaction can explain the trace.
It produces an error, alerting the designer to a protocol violation.

\begin{figure*}
  \centering
  \footnotesize
  \setlength{\arraycolsep}{1.5pt}

  \begin{minipage}[t]{0.58\textwidth}
  \centering
  \[
  \begin{array}{rlll}
  \langle stmt\rangle ::= & \langle in\_port\rangle\texttt{ := }\langle bv\_expr\rangle & & \textit{assign port} \\
  \mid & \langle in\_port\rangle\texttt{ := X} & & \textit{DontCare assignment} \\
  \mid & \texttt{step(}k\texttt{)} & & \textit{advance $k$ clock cycles} \\
  \mid & \texttt{fork()} & & \textit{start concurrent transaction} \\
  \mid & \texttt{assert\_eq(}\langle bv\_expr\rangle\texttt{, }\langle bv\_expr\rangle\texttt{)} & & \textit{check equality} \\
  \mid & \langle stmt\rangle\texttt{ ; }\langle stmt\rangle & & \textit{sequential composition} \\
  \mid & \texttt{if }\langle bv\_expr\rangle\texttt{ \{ }\langle stmt\rangle\texttt{ \} }\texttt{else}\texttt{ \{ }\langle stmt\rangle\texttt{ \}} & & \textit{conditional branch} \\
  \mid & \texttt{while }\langle bv\_expr\rangle\texttt{ \{ }\langle stmt\rangle\texttt{ \}} & & \textit{conditional loop} \\
  \mid & \texttt{repeat } \langle uint\_param \rangle \texttt{ iterations \{ }\langle stmt\rangle\texttt{ \}} & & \textit{$k$-iteration loop} \\
  \mid & \texttt{for }\langle x\rangle\texttt{ in }\langle seq\_param\rangle\texttt{ \{ }\langle stmt\rangle\texttt{ \}} & & \textit{iterate over sequence} \\[4pt]
  \langle prot\rangle ::= & \texttt{prot name<DUT: }\langle RTL\_module\rangle\texttt{>(} \\
  & \epsilon\mid\langle param\rangle(\texttt{,}\langle param\rangle)^*\texttt{) \{ }\langle stmt\rangle\texttt{ \}} & & \textit{protocol definition}
\end{array}
  \]
  \end{minipage}\hspace*{-14pt}
  \begin{minipage}[t]{0.43\textwidth}
  \centering
  \[
  \begin{array}{rlll}
  \langle bv\_expr\rangle ::= & \langle b\rangle & & \textit{bit-vector constant} \\
  \mid & \langle port\rangle & & \textit{I/O port on the DUT} \\
  \mid & \langle bv\_param\rangle & & \textit{bit-vector parameter} \\
  \mid & \langle bv\_expr\rangle\mathbin{\oplus}\langle bv\_expr\rangle & & \textit{$\oplus\in\{\texttt{==},\texttt{\&\&},\texttt{+}\}$} \\
  \mid & \langle bv\_expr\rangle\texttt{ \#\# }\langle bv\_expr\rangle & & \textit{concatenation} \\
  \mid & \texttt{!}\langle bv\_expr\rangle & & \textit{logical negation} \\
  \mid & \langle bv\_expr\rangle\texttt{[}msb\texttt{:}lsb\texttt{]} & & \textit{bit-slice} \\
  \mid & \texttt{is\_last()} & & \textit{last iteration? (repeat/for-in)} \\
  \mid & \texttt{iter\_count()} & & \textit{iteration count (repeat/for-in)} \\
  \langle param\rangle ::= & \langle bv\_param\rangle\texttt{: u}\langle n\rangle & & \textit{bit-vector parameter} \\
  \mid & \langle uint\_param\rangle\texttt{: uint} & & \textit{integer parameter} \\
  \mid & \langle seq\_param\rangle\texttt{: [u}\langle n\rangle\texttt{]} & & \textit{sequence parameter}
  \end{array}
  \]

  \end{minipage}

  \caption{Syntax for the \lang{} DSL.}
  \label{fig:paso-language-summary}
\end{figure*}

\section{The \lang{} Language}
\label{sec:the_paso_language}

In \lang{}, the user defines protocols that can take arguments and operate over the input and output ports of a design under test (DUT).
Protocols are analogous to functions or tasks in traditional testbench languages.
Multiple protocols can be defined over the same DUT (e.g. \code|add| and \code|mul| protocols for an ALU, \code|read| and \code|write| transactions for a memory bus).
Figure~\ref{fig:paso-language-summary} shows all statements and expressions that can make up the body of a protocol.
They are modeled on imperative testbench frameworks like ChiselTest, Cocotb and a certain style of traditional Verilog testbenches~\cite{system_verilog_spec, cocotb, chiseltest}.

\subsection{Single Protocol Semantics}

We first discuss how a single protocol executes in isolation.
When a protocol execution is launched, it receives immutable transaction arguments (in the \code|add| protocol shown in Figure~\ref{fig:adder_protocols} the values of $a$, $b$, and $s$) and interacts with the DUT (Figures~\ref{fig:combinational_adder} and~\ref{fig:pipelined_adder}) through its input and output ports. 

\paragraph{Design under test}
To define the semantics of \lang{} as it interacts with a DUT, we model the DUT as a fully deterministic transition system with possible inputs $I$, states $S$, and outputs $O$.
The values on the output ports of the DUT can be computed from the current input and state values: $I \times S \rightarrow O$.
When the \code|step| statement is reached, the internal state of the DUT is similarly updated to a next state: $I \times S \rightarrow S$.
All input and output ports of the DUT carry a valid 2-state value at all times.
We also assume access to combinational static cone of influence information.
In particular, we need to know which input ports may directly influence which output ports.
This information is easily available, as it is also used to find combinational loops in RTL designs~\cite{wire_sorts}.
This model of our DUT is general enough to fit software simulation, FPGA emulation, and formal verification.
We chose to forgo any X-propagation through the DUT because it is not uniformly supported \cite{turpin_2003, sutherland_2013}, and the default SystemVerilog semantics are neither sound nor complete \cite{loow_oopsla25, choi_oopsla2025}.

A protocol manipulates the input values ($I$) through assignments to input ports of the DUT.
Each input port acts like a state variable that maintains its value until reassigned.
An input port can be assigned a \texttt{DontCare} value (\code|X|) to indicate that its value is not relevant to the execution of the protocol.
At the start of a protocol execution and after the final \texttt{step}, all inputs are implicitly assigned \texttt{DontCare}.
This means that we do not make any assumptions about the input values when a protocol starts.
On the flipside, the protocol fully relinquishes control over all input ports as it finishes.

\paragraph{Transient values (glitches)}
In a synchronous design, each wire converges to a final value before the end of a clock cycle.
In the hardware implementation, only this final value matters.
However, in a testbench written in an imperative language, it is normally rather easy to observe transient values.
Take the combinational adder from Figure~\ref{fig:combinational_adder}:
A protocol might evaluate an expression containing \texttt{DUT.s} and thus observe that output. Now, if either of the inputs is reassigned in the same cycle, that would have an immediate effect on \texttt{DUT.s} and thus the first observation would have been transient.
Transient values do not show up in waveform outputs and cannot be reliably captured in real hardware.
Thus, they are a frequent source of hard-to-find bugs in SystemVerilog testbenches \cite{loow_oopsla25}. 

\paragraph{Unique value restriction}
We restrict protocols to only observing a single unique value for each output port in a single step.
To do so, a protocol is not allowed to assign to an input that may change the value of an output port that has been observed in the same step.
This restriction can easily be implemented in a sound, but slightly imprecise way: when an input port is assigned, we compute the set of ports in the combinational static cone of influence and the set of ports already observed in the current step. If the intersection is non-empty, our unique value restriction may be violated.
We implemented this check in the \lang{} interpreter, and thus far have not observed any false positives.

\paragraph{Control flow}
\lang{} supports conditional branches and while loops, which follow standard imperative semantics and can use DUT outputs to determine future control flow.
\code|repeat| and \code|for-in| loops allow for a variable number of repetitions controlled by function arguments. 

\begin{figure}[b]
  \centering
\begin{lstlisting}[language=paso,escapechar=!,aboveskip=0pt,belowskip=0pt]
prot add<DUT: Adder>(a: u32, b: u32, s: u32) {
  DUT.a := a; DUT.b := b; 
  assert_eq(DUT.s, s);    
  step(); fork();
  // Missing X assignments to DUT.a and DUT.b 
  step(); ...
}
\end{lstlisting}
  \vspace{-3mm}
  \includegraphics[width=0.8\linewidth]{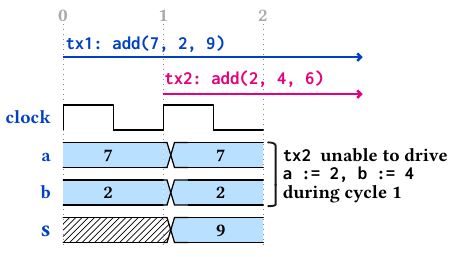}
  \vspace{-3mm}
  \caption{
    \label{fig:naive_pipelined_adder}
    An erroneous version of the pipelined adder protocol
    from Figure \ref{fig:pipelined_adder},
    executed with transactions \texttt{tx1}, \texttt{tx2}.
  }
\end{figure}

\subsection{Concurrent Protocol Semantics}
\label{sec:concurrent_protocol_semantics}

RTL modules generally describe reactive systems that may execute for an essentially unbounded number of cycles.
In the same spirit, we mandate that a DUT that abides by a given \lang{} specification shall always be executing at least one transaction.
This is achieved by launching a new transaction as soon as the previous one finishes.
However, we often need to overlap the transaction executions, such as for pipelined circuits as shown in $\S$\ref{sec:pipelined_adder}.
To permit concurrent transactions, a call to \code|fork| launches exactly one new protocol execution. A single protocol execution can only fork once.
If a protocol finishes without having called \code|fork|, a new protocol execution is launched after the final \code|step|, leading to sequential rather than parallel composition.

Protocols can only fork after at least one call to \code|step|. These restrictions ensure that the reconstructor never needs to consider an unbounded number of executions spawned in a single cycle; in particular, disallowing \code|fork| prior to any calls to \code|step| guarantees that only one new protocol execution can spawn each cycle. Consequently, the set of possible combinations of live executions remains finite and tractable for the reconstructor to explore, as described in \S\ref{sec:path-exploration}.

We now describe how \lang's concurrency semantics avoid race conditions, which are a common debugging nightmare in traditional testbench setups.

\paragraph{Synchronization}
When we have multiple concurrent transactions, we want the outcome to be independent of the order in which they are executed.
In \lang{}, the only shared variables are the inputs and the state of the DUT.
The DUT state can only be updated by \code|step| which we treat as a synchronization barrier:
when a transaction reaches a \code|step| statement, its execution is suspended until all other transactions have also reached a \code|step|.
Only then is the DUT state updated to the next state.
Thus, each transaction sees exactly one state update for every \code|step|.

To synchronize inputs, we mandate that all final input assignments across concurrent transactions are compatible.
Assignments to an input port are compatible if they specify the same value, or if the other value is \code|X|. Thus, two threads may assign, e.g., 1 and 1, \code|X| and 1 or \code|X| and \code|X|. However, they may not assign, e.g., 1 and 2 to the same DUT input port.
In Figure~\ref{fig:naive_pipelined_adder}, a faulty \texttt{add} protocol fails to relinquish control over the input ports in its second cycle, leading to conflicting assignments across two invocations with distinct parameters.

\paragraph{Unique value restriction}
To achieve true isolation between concurrent transactions and thus race freedom, we need to strengthen the unique value restriction to handle the parallel case.
Recall we forbid assigning to an input port if it may change the value of an output port that has already been observed in the same cycle. 
In the single-protocol case, this restriction makes an observed output stable for the remainder of the step: the current transaction cannot change any contributing input, and no other transaction is running to do so. With concurrent transactions, however, another transaction may replace a port assigned to \texttt{X} with a concrete value without triggering a conflict, mutating the input port, and thereby, the previously observed output port.
For example, suppose \code|DUT.in| combinationally influences \code|DUT.out|, and transaction A sets \code|DUT.in := X| and observes \code|DUT.out| in the same cycle. If transaction B concurrently assigns a value to \code|DUT.in|, A's observation of \code|DUT.out| may be transient. If all contributing inputs have concrete values, this cannot happen: a different concrete assignment by a concurrent transaction triggers a conflict, while a \texttt{DontCare} assignment leaves the concrete value unchanged. Therefore, we additionally require that before observing an output port, a transaction must assign concrete (non-\code|X|) values in the current cycle to every input port that can affect that output.

\section{Reconstructing Transaction Traces}
\label{sec:monitor}

So far, we have mostly discussed the ``forward'' semantics, i.e., how \lang{} protocols are executed to test or control a DUT.
However, engineers routinely write similar drivers in any of the available testbench frameworks.
What sets \lang{} apart as a true specification language is the fact that we can use the same protocol description for analyzing waveforms.

The \lang{} \emph{\monitorname{}} recognizes transactions according to a protocol specification.
It analyzes a signal trace and constructs a \emph{transaction trace} that is consistent with the waveform.
Crucially, the \monitorname{} does not need to execute the DUT; it only uses a waveform from RTL simulation.
If no transactions could have resulted in the waveform trace based on the input protocol definitions, the \monitorname{} reports an error.

\subsection{Inspiration from Dynamic Symbolic Execution}

Our reconstruction algorithm is inspired by dynamic symbolic execution (DSE)~\cite{clarke_1976, king_1976,dart:pldi05, exe:ccs06}.
A DSE engine tries to find inputs to a program that trigger an assertion violation.
This mechanism can be adapted to reverse a software function, like computing a value for the symbolic input \code|a| that gets a pure function \code|foo| to return \code|5|:

\begin{lstlisting}
int a = input(); assert(foo(a) != 5);
\end{lstlisting}

To explain our reconstruction algorithm, consider the hypothetical \lang{} interpreter as the function being reversed. This interpreter takes a transaction trace as its symbolic input and produces a signal trace as its output. It also has access to the user-defined protocol specifications and the DUT, much like how \code|foo| has access to its arguments and computes a return value.

Our \monitorname{} differs from conventional DSE in two ways. First, note how we inverted the assertion in our example; this is necessary because DSE engines search for a path that \emph{violates} an assertion. However, our goal is not to find an execution path that violates an assertion. Instead, the \monitorname{} searches for an execution path and transaction arguments that produce the given signal trace without violating any \code|assert_eq| statements in the \lang{} protocols.

Second, the \lang{} interpreter also differs from the pure function \code|foo| in that it continuously interacts with the DUT while executing transactions. Assertions are therefore checked throughout execution, rather than against a final return value. The values assigned to DUT inputs and the values observed on DUT outputs are determined by the provided signal trace. Each assignment or observation can consequently reveal an inconsistency with the trace, allowing the \monitorname{} to prune infeasible paths early. If a protocol assigns 0 to \code|DUT.op| or asserts \code|assert_eq(DUT.op, 0)|\footnote{\label{note:assign_assert_caveat}During reconstruction, assignments, and equality assertion both impose equality constraints and thus appear the same. However, the distinction matters when forward executing a protocol to drive a DUT. Additionally, the unique value restriction discussed in Section~\ref{sec:the_paso_language} requires input ports to be assigned before their value can be observed.}, but the trace shows that \code|DUT.op| is \code|1| at the current step, we know that this path is infeasible. In contrast, a conventional DSE engine may need to explore all paths through \code|foo| until the function exits before determining whether it satisfies the target condition.

DSE engines are often plagued by scalability issues due to path explosion~\cite{baldoni_et_al_2018, cadar_sen_cacm_2013, schmemmel_et_al_cav_2020} and the NP-hardness of solving path constraints~\cite{cook_stoc1971,de_moura_bjorner_cacm_2011}.
However, for \lang{} protocol specifications, we generally do not face these issues.
The DSL was carefully designed to minimize branching on symbolic values, thus limiting the number of program paths that need to be tracked.
A breadth-first path exploration strategy allows us to process signal traces as a stream, keeping memory requirements low.
Finally, all constraints that we need to consider are simple (in-) equalities to a constant, which can be solved in constant time.

\begin{figure*}
  \begin{minipage}[t]{0.50\textwidth}
    \begin{minipage}{.45\linewidth}
\begin{lstlisting}[language=paso,escapechar=|,aboveskip=0pt,belowskip=0pt,numbers=left,numberstyle=\footnotesize\color{pasoGray}]
prot add<DUT>(a, b, s) {
  DUT.a := a; DUT.b := b;
  |\textcolor{pasoRed}{DUT.op := 0;}|
  step();
  DUT.a := X; DUT.b := X;
  DUT.op := X;
  assert_eq(DUT.s, s);
  fork(); step();
}
\end{lstlisting}
    \end{minipage}%
    \hspace{3mm}\vline\hspace{5mm}%
    \begin{minipage}{.45\linewidth}
\begin{lstlisting}[language=paso,escapechar=|,aboveskip=0pt,belowskip=0pt]
prot sub<DUT>(a, b, d) {
  DUT.a := a; DUT.b := b;
  |\textcolor{pasoRed}{DUT.op := 1;}|
  step();
  DUT.a := X; DUT.b := X;
  DUT.op := X;
  assert_eq(DUT.s, d);
  fork(); step();
}
\end{lstlisting}
    \end{minipage}
  \end{minipage}%
  \hfill%
  \begin{minipage}[c]{0.46\textwidth}
    \centering
    \includegraphics[width=0.8\linewidth]{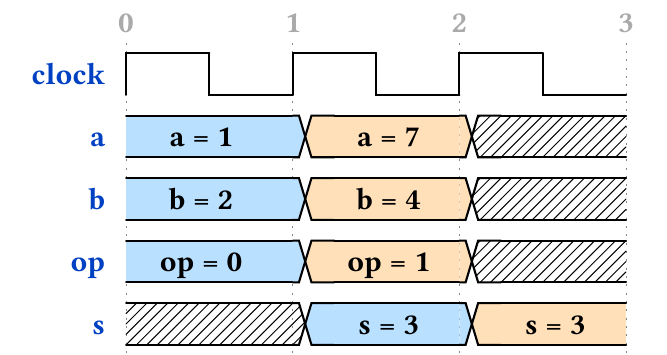}
  \end{minipage}

  \vspace{3pt}

  \begin{subfigure}[t]{0.48\textwidth}
    \centering
    \begin{tikzpicture}[thr/.style={draw, rounded corners, minimum
        width=2.2cm, minimum height=0.45cm, font=\small, align=center},
        root/.style={draw, circle, minimum size=0.4cm,
      font=\footnotesize, fill=pasoGray!30}]
      \node[root] (r) at (0,0) {$\varnothing$};
      \node[thr, fill=pasoRed!25, right=0.5cm of r, yshift=0.55cm] (s)
      {\code|sub|($a,b,d$)\\{\color{pasoGray}\{\}}};
      \node[thr, fill=pasoBlue!20, right=0.5cm of r, yshift=-0.55cm] (a)
      {\code|add|($a,b,s$)\\{\color{pasoGray}\{\}}};
      \draw[->, pasoGray] (r) -- (s.west);
      \draw[->, pasoGray] (r) -- (a.west);
    \end{tikzpicture}
    \subcaption{The \monitorname{} begins by exploring one execution path
    per protocol with no constraints.}
  \end{subfigure}%
  \hfill%
  \begin{subfigure}[t]{0.48\textwidth}
    \centering
    \begin{tikzpicture}[thr/.style={draw, rounded corners, minimum
        width=2.2cm, minimum height=0.45cm, font=\small, align=center},
        root/.style={draw, circle, minimum size=0.4cm,
      font=\footnotesize, fill=pasoGray!30}]
      \node[root] (r) at (0,0) {$\varnothing$};
      \node[thr, fill=pasoRed!25, right=0.5cm of r, yshift=0.6cm] (s)
      {\code|sub| $\times$\\\footnotesize $\{a{=}1,b{=}2,1{=}0\}$};
      \node[thr, fill=pasoBlue!20, right=0.5cm of r, yshift=-0.6cm] (a)
      {\code|add|(1,2,$s$)\\\footnotesize$\{a{=}1,b{=}2,0{=}0\}$};
      \draw[->, pasoGray] (r) -- (s.west);
      \draw[->, pasoGray] (r) -- (a.west);
    \end{tikzpicture}
    \subcaption{\textbf{Step 1}:
    Similar assignments in \code|add| and \code|sub| constrain \code|a| and \code|b| to concrete values. The path executing \code|sub| is pruned because, in this step, \code|DUT.op := 1| implies $ 0 = 1$ which is false.}
  \end{subfigure}

  \vspace{3pt}

  \begin{subfigure}[t]{0.48\textwidth}
    \centering
    \begin{tikzpicture}[thr/.style={draw, rounded corners, minimum
        width=2.2cm, minimum height=0.45cm, font=\small, align=center},
        root/.style={draw, circle, minimum size=0.4cm,
      font=\footnotesize, fill=pasoGray!30}]
      \node[root] (r) at (0,0) {$\varnothing$};
      \node[thr, fill=pasoRed!25, right=0.4cm of r, yshift=0.6cm] (s0)
      {\code|sub| $\times$};
      \node[thr, fill=pasoGreen!25, right=0.4cm of r, yshift=-0.6cm]
      (a0) {\code|add|(1,2,3) \checkmark};
      \node[thr, fill=pasoRed!25, right=0.5cm of a0, yshift=0.5cm] (a1)
      {\code|add| $\times$\\\footnotesize $\{a{=}7,b{=}4,0{=}1\}$};
      \node[thr, fill=pasoOrange!20, right=0.5cm of a0, yshift=-0.5cm]
      (s1) {\code|sub|(7,4,$d$)\\\footnotesize\{$a{=}7,b{=}4,1{=}1$\}};
      \draw[->, pasoGray] (r) -- (s0.west);
      \draw[->, pasoGray] (r) -- (a0.west);
      \draw[->, pasoGray] (a0) -- node[above, font=\footnotesize,
      text=pasoGray]{fork} (a1.west);
      \draw[->, pasoGray] (a0) -- (s1.west);
    \end{tikzpicture}
    \subcaption{\textbf{Step 2}: The first \code|add| completes with $s=3$. It also calls \code|fork()| which leads to a split in the execution path.
    The path executing a new \code|add| is pruned because, in this step, \code|DUT.op := 0| implies $1 = 0$.}
  \end{subfigure}%
  \hfill%
  \begin{subfigure}[t]{0.48\textwidth}
    \centering
    \begin{tikzpicture}[thr/.style={draw, rounded corners, minimum
        width=1.9cm, minimum height=0.45cm, font=\small, align=center},
        root/.style={draw, circle, minimum size=0.4cm,
      font=\footnotesize, fill=pasoGray!30}]
      \node[root] (r) at (0,0) {$\varnothing$};
      \node[thr, fill=pasoRed!25, right=0.4cm of r, yshift=0.6cm] (s0)
      {\code|sub| $\times$};
      \node[thr, fill=pasoGreen!25, right=0.4cm of r, yshift=-0.6cm]
      (a0) {\code|add|(1,2,3) \checkmark};
      \node[thr, fill=pasoRed!25, right=0.5cm of a0, yshift=0.4cm] (a1)
      {\code|add| $\times$};
      \node[thr, fill=pasoGreen!25, right=0.5cm of a0, yshift=-0.4cm]
      (s1) {\code|sub|(7,4,3) \checkmark};
      \draw[->, pasoGray] (r) -- (s0.west);
      \draw[->, pasoGray] (r) -- (a0.west);
      \draw[->, pasoGray] (a0) -- node[above, font=\footnotesize,
      text=pasoGray]{fork} (a1.west);
      \draw[->, pasoGray] (a0) -- (s1.west);
    \end{tikzpicture}
    \subcaption{
      The \monitorname{} terminates, with the final inferred
      transaction trace being \code|add(1,2,3); sub(7,4,3)|.
    }
  \end{subfigure}
  \vspace{-2pt}
  \caption{
    \label{fig:alu_waveform}\label{fig:reconstructor_walkthrough}
    Example \lang{} protocol specification and waveform for a pipelined ALU.
    The two protocols assign different values to \code|DUT.op| during the first step of their execution. This enables the \monitorname{} to distinguish them.
  }
\end{figure*}

\subsection{Path Exploration}
\label{sec:path-exploration}

A \emph{program path} in DSE corresponds to an incomplete transaction trace in our \monitorname{}.
We start with an empty trace and an implicit \code|fork| statement that starts the first transaction.
At this point, we do not know which protocol executes first, and thus we treat the \code|fork| statement like a symbolic $n$-way branch, where $n$ is the number of available protocol definitions.
Each of the $n$ paths starts a different protocol with unknown, i.e., symbolic, arguments. Subsequent \code|fork| statements are handled in the same way, creating $n$ branches per live path for the possible new concurrent transactions.

The \monitorname{} then executes each path until it either reaches the end of a clock cycle or a constraint violation occurs (we discuss the exact nature of constraints in $\S\ref{sec:reconstructor_constraints}$.) 
A constraint violation indicates that a path is infeasible, meaning that it cannot explain the given signal trace. As long as there are still other paths executing, the search continues. In practice, the statements inside different protocols often impose incompatible constraints on the DUT ports, causing all paths but one to be pruned quickly. Just as \code|step| acts as a synchronization barrier in the forward semantics of \lang{}, we pause paths that reach the end of a step until all other active paths have also finished.
By keeping all paths synchronized, we can stream signal values cycle-by-cycle without ever having to backtrack.
Figure~\ref{fig:reconstructor_walkthrough} illustrates this process with an example. 

In \lang{}, guards to \code|if| statements and \code|while| loops may not involve any protocol parameters.
Only DUT ports and constants are allowed to appear in guards: since the \monitorname{} has access to the full signal trace, DUT ports always evaluate to a known value.
Thus, \code|if| and \code|while| branches can always be fully resolved and do not lead to new program paths that need to be explored.

Argument-dependent control flow is enabled by the specialized \code|repeat| and \code|for-in| loops.
The number of iterations for these loops depends on the value of an integer parameter or the length of a sequence argument respectively.
For these loops, we need to check the feasibility of exiting or staying in the loop for every iteration and potentially explore both resulting paths.

\subsection{Constraints}
\label{sec:reconstructor_constraints}

\lang{} has no local variables, and protocol parameters are immutable.
Every time the \monitorname{} launches a new protocol on an execution path, it creates fresh symbolic variables to represent the transaction arguments.
For each parameter, the \monitorname{} maintains the least information needed to represent its current set of possible values. For bit-vector parameters, it tracks which bits are known and their values, initially with no known bits. For \code|uint| parameters, it tracks a lower bound, initially zero. For sequence parameters, it tracks a minimum length and lazily generates fresh bit-vector symbols as that length increases.

Bit-wise equality constraints arise from assignments and equality assertions. For assignments, the left-hand side is always a DUT input port, which has a concrete value from the signal trace.
For \code|assert_eq|, we require that one side of the equality is either a DUT port or a bit-vector literal. This restriction has not limited our desired expressivity in practice. 
The other expression is either a bit-vector type protocol parameter, a constant, or a concatenation of constants and protocol parameter bits.
If we split the equality of two $n$-bit expressions into $n$ 1-bit equalities, we are left with either simple constant equalities ($1=1$ or $1=0$) or an equality of an argument bit to a constant (\code|a[0] = 1|). Constant equalities are trivially false or true. For argument equalities, we first check whether the value of the bit is already known. If it is unknown, we update the value with the newly learned bit. If the bit is known, we compare with the current value, discarding the path if equality fails.

Note that assignments can be overwritten in the same step, which can mask their effect. That is, the value of the first assignment is transient and does not appear in the signal trace. We deal with this fact by delaying the evaluation of assignments until the \code|step| barrier, allowing us to only consider the \emph{final} assignment to each port.

The one use for integer-type protocol parameters is to control the number of iterations of a \code|repeat| loop.
The first time we encounter a \code|repeat| loop with parameter $n$, we only know that $n \geq 0$.
At this point, the \monitorname{} creates two paths: one in which
$n=0$ and the loop is skipped, and another in which $n\geq 1$ and the
loop body is entered. After executing the loop body, it branches again,
now considering the cases $n=1$ and $n\geq 1$.
Any reasonable \code|repeat| loop we have encountered contains at least one \code|step| and thus we only branch once per clock cycle.
As with branching due to fork, the statements inside the loop body
and those following the loop often impose incompatible constraints leading to rapid pruning.
If the loop body contains an \code|is_last()| expression, the \monitorname{} branches early, since the loop body now depends on whether this is the last loop iteration or not.

Sequence-type parameters for \code|for-in| loops are treated almost the same, except we impose constraints on the sequence length.
When the minimum length of a sequence $s$ is incremented, for example, when going from \code|len(s)|~$\geq 0$ to \code|len(s)|~$\geq 1$, we create a fresh symbolic bit-vector value to represent \code|s[0]|.
This symbolic value is treated the same as the scalar bit-vector parameters described earlier.

\begin{table}[b]
\caption{
We wrote \lang{} protocol specifications for 12 open-source RTL modules.
All protocols have been tested with our interpreter and our \monitorname{}.
The table shows the lines of code (LoC) and number of \code|prot| definitions.
}
\label{fig:other_duts_table}
\footnotesize
\setlength{\tabcolsep}{5pt}
\raggedright
\begin{tabular}{lrr}
\toprule
\textbf{DUT Description} &
\textbf{LoC} &
\textbf{\# \code|prot|}
\\
\midrule
AES-128 Encryption Core \cite{tiny_aes} & 30 & 3  \\
APB Manager \cite{raj_apb} & 58 & 4  \\
APB SRAM Subordinate \cite{amba_apb_sram} & 58 & 4  \\
AXI-Stream Adapter \cite{forencich_verilog_axis} & 145 & 3  \\
AXI-Stream Asynchronous FIFO \cite{forencich_verilog_axis} & 75 & 6  \\
AXI-Stream Manager \cite{zipcpu_xlnxstream2018_3} & 23 & 2  \\
BaseJump STL FIFO \cite{basejump_stl} & 30 & 4  \\
Multiplier (PicoRV32 RISC-V Core) \cite{picorv32} & 33 & 3  \\
Register file (Engine-V RISC-V Core) \cite{engine_v} & 23 & 1  \\
Register file (SERV RISC-V Core) \cite{serv} & 190 & 2  \\
Wishbone SRAM \cite{antmicro_2022_wishbone} & 291 & 17  \\ 
Wishbone FIFO \cite{antmicro_2022_wishbone} & 291 & 17  \\ 
\bottomrule
\end{tabular}
\end{table}

\section{Case Studies}
\label{sec:case_studies}

We demonstrate \lang{} using case studies that analyze the protocols in existing, open-source RTL designs.
We answer the following research questions (RQs):
\begin{enumerate}[label=RQ\arabic*:]
  \item Is \lang{} expressive enough to describe real-world on-chip communication protocols?
  \item How does the \monitorname{}'s performance scale with long waveforms and increasing numbers of protocol definitions?
  \item To what extent does our \monitorname{} help identify protocol violation bugs in real-world RTL designs?
\end{enumerate}
To address RQ1, we write \lang{} protocols to describe complex features in the Wishbone ($\S\ref{sec:wishbone}$) and AXI-Stream ($\S\ref{sec:protocol_bugs_case_studies}$) protocols. 
We also describe other RTL designs we have successfully driven and monitored using \lang{}.
To address RQ2, we measure the performance of the \monitorname{} on long traces obtained from the 
testbenches for real-world Wishbone RTL designs. Each trace contains over 75,000 cycles, with the 
longest waveform containing over 31 million cycles ($\S\ref{sec:reconstructor_performance}$). 
To address RQ3, we demonstrate how the \monitorname{} can identify violations of the AXI-Stream specification 
in real-world RTL implementations, focusing on a dataset of RTL design bugs published in prior work \cite{ma_et_al_asplos_2022} ($\S\ref{sec:protocol_bugs_case_studies}$).

\subsection{Expressivity}

To evaluate \lang{}'s expressivity, we wrote a range of \lang{} protocols that allow us to both drive and monitor
various real-world RTL designs. Table~\ref{fig:other_duts_table} lists the projects, all obtained from public GitHub repositories, and statistics about the protocol.
These protocols include components from various RISC-V cores and implementations of standardized protocols, e.g. AXI-Stream, APB and Wishbone.
One notable DUT is a pipelined AES-128 encryption core \cite{tiny_aes}, which has a known latency of 20 cycles. This particular DUT 
is a latency-\emph{sensitive} interface (unlike other latency-\emph{insensitive} protocols, e.g., AXI-Stream),
indicating that both latency-insensitive and sensitive interfaces can be expressed in \lang{}. 
Another notable DUT is a register file from the SERV RISC-V core \cite{serv}, in which 32-bit data is transmitted 1 bit per cycle during 
read and write transactions. The corresponding \lang{} protocol uses bit-slicing operations extensively, 
exercising the \monitorname{} functionality for tracking known bits of bit-vectors.

\subsubsection{Wishbone}
\label{sec:wishbone}
To push the limits of what we can express with \lang{}, we wrote a fairly complete specification of the Wishbone protocol.
Wishbone is an open-source communication bus for on-chip interconnects, developed by the OpenCores initiative \cite{Wishbone:2010}. Beyond individual reads and writes, Wishbone contains many complex features, such as various burst modes.

We wrote \lang{} protocols to describe these Wishbone features.
To gain confidence that our \lang{} protocols correctly specify the behavior of different Wishbone features,  we manually constructed signal traces depicting transactions that exercise various aspects in the Wishbone B4 specification \cite{Wishbone:2010}.
Then, we ran the \monitorname{} on these signal traces, manually inspecting the resultant transaction trace to ensure they are consistent with the Wishbone specification.

We then used our Wishbone specification on waveform traces collected from three different open-source projects: a Wishbone SRAM and FIFO from Antmico~\cite{antmicro_wb_burst_benchmark}, an Ethernet IP from OpenCores~\cite{ethmac}, and a Wishbone streaming IP~\cite{wb_streamer}. While these projects did use different pin names and address port sizes, we were able to use the exact same Wishbone protocol module pin renaming on all of them.
Together, the signal traces we collected covered normal reads and writes, as well as all burst modes supported by Wishbone.
We did discover one significant bug in the Antmicro SRAM, for which our \monitorname{} reported an error, which we describe in more detail in Section~\ref{sec:protocol_bugs_case_studies}.

\subsection{Reconstructor Performance}
\label{sec:reconstructor_performance}
Next, to evaluate the performance of the \monitorname{},
we picked two open-source Wishbone implementations on GitHub whose testbenches 
produced long waveforms (over 75,000 cycles) \cite{ethmac, wb_streamer}. We wrote 17 protocols
specifying the behavior of single, block, burst transfers 
in different scenarios.
Appendix~\ref{sec:wishbone-appendix} lists the full \lang{} code for these protocols.

\begin{table}[b]
\caption{
  Wall-clock time of the \monitorname{} on long Wishbone (WB) waveforms obtained from GitHub ~\cite{ethmac, wb_streamer}.
  WB transactions (abbreviated Tx.) denote reads and writes. The benchmarks are an OpenCores WB Ethernet core and four configurations of a FuseSoc WB data streaming core.
}
\label{fig:wishbone_bi_benchmark_table}
\centering
\footnotesize
\setlength{\tabcolsep}{2pt}
\begin{tabular}{L{2.3cm}rrrrr}
\toprule
\multirow{4}{*}{\small\textbf{Benchmark:}}
& & \multicolumn{4}{c}{\textbf{Wishbone Stream}} \\
\cmidrule(l){3-6}
& & \multicolumn{2}{c}{\textbf{Reader}}& \multicolumn{2}{c}{\textbf{Writer}} \\
\cmidrule(l){3-4}\cmidrule(l){5-6}

&
\textbf{Ethernet} &
\textbf{Control} &
\textbf{Data} &
\textbf{Control} &
\textbf{Data}
\\
\midrule
\textbf{Mean Runtime}    & 11min &  10.69s &  10.82s &  1.53s &  4.64s \\
\textbf{Total Cycles}    & 31,663,327 & 201,903 & 201,903 & 91,165 & 91,165 \\
\textbf{WB Transactions} &    344,744 &   4,473 &   5,000 &  4,377 &  5,000 \\
\textbf{Idle Cycles}     &       95\% &    89\% &    95\% &   26\% &   79\% \\
\textbf{Mean Tx. Length} &       4.93 &    5.05 &    2.00 &  15.49 &   2.00 \\
\bottomrule
\end{tabular}
\end{table}

We ran the \monitorname{} against these waveforms on a server with dual Intel Xeon E5-2620v4 CPUs with 8 cores.
We measured wall-clock time using the Hyperfine benchmarking tool \cite{hyperfine_2023}, averaged over ten runs.

\begin{figure}
\centering
\includegraphics{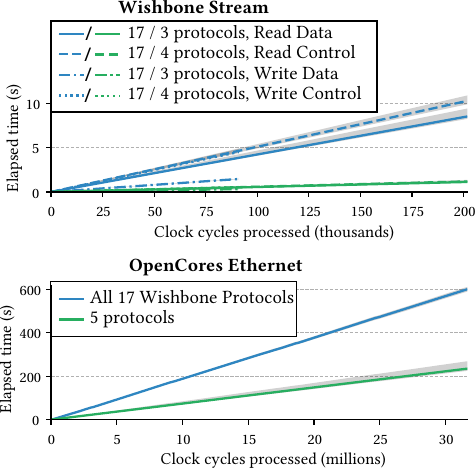}
\caption{
\label{fig:reconstructor_ethmac_timing_plot}
The \monitorname{} wall-clock time scales with the number of clock cycles processed on various long Wishbone waveforms.
The blue lines show timing when we include our full Wishbone specification with 17 protocol definitions. For the green lines, we only included the protocols that actually appear in the waveform.
We plot the mean over 10 runs, with minimums and maximums shaded in grey.  
}
\end{figure}

Table~\ref{fig:wishbone_bi_benchmark_table} contains the wall-clock time and other statistics for reconstruction.
We observe that the \monitorname{} performance scales to long waveforms, with the longest waveform (over 31 million cycles)
processed in approximately 11 minutes. In Figure~\ref{fig:reconstructor_ethmac_timing_plot}, we compare the wall-clock 
time of the \monitorname{} on these long Wishbone waveforms with two sets of protocols: 17 protocols describing different Wishbone features, 
and a subset which only contains the transactions that occur in the inferred transaction trace.
The \monitorname{} runtime 
scales proportionally with the number of clock cycles in the waveform and the number of protocol specifications.

\subsection{Uncovering Protocol Violation Bugs}
\label{sec:protocol_bugs_case_studies}

Our case studies using the \lang{} \monitorname{} identified known and unknown bugs in hardware protocol implementations.

\subsubsection{Wishbone Burst Mode Bug} 
In Wishbone's burst mode, within the same burst transfer, addresses can be incremented automatically in a \emph{wrapped} fashion.
In this mode, the address is first incremented ordinarily. Then, given a user-selected \emph{wrap size} of $n$,
the $\log_2(n)$ least significant bits (LSBs) of the resultant address are computed modulo $n$ \cite{Wishbone:2010}.
Wrapped increments are useful for situations such as reading from a cache line, 
where the address must be aligned with cache line boundaries. 

We found a bug in the burst mode implementation of a Wishbone SRAM designed by Antmicro \cite{antmicro_wb_bug}. When we ran our \monitorname{} with our common Wishbone specification on the waveform output of the provided testbenches. On certain burst sizes and address offsets, our \monitorname{} would report an error, which we show in Figure ~\ref{fig:reconstructor_wishbone_burst_error_msg}.
Looking at the code, we believe that the Antmicro engineers misunderstood how exactly addresses are calculated in the wrapping burst mode. The official Wishbone specification is very vague on this issue, and we were only able to learn the correct behavior from looking at a different Wishbone implementation.
The SRAM design from Antmicro was merged into the upstream Litex project \cite{litex_wishbone}. However, it appears that this particular wrap mode is never used, and thus it is unlikely that any Litex user has encountered this bug.

\begin{figure}[b]
{\catcode`\^=12
\begin{lstlisting}[aboveskip=0pt,belowskip=0pt,basicstyle=\ttfamily\footnotesize,moredelim={[is][\color{pasoRed}]{@}{@}}]
@error@: [read_burst_increment_wrap] transaction failed at 500ns!
DUT.addr := start_addr[31:5] ## (start_addr[4:2] + iter_count()) 
              ## start_addr[1:0];
^^^^^^^^
@expected DUT.addr = 0x04000000 but got 0x04000008@
\end{lstlisting}
}
\caption{
  \label{fig:reconstructor_wishbone_burst_error_msg}
  Reconstructor error indicating a violation 
  of the Wishbone spec for wrapped addresses in burst mode.
}
\end{figure}

\begin{table}[b]
\caption{
  Protocol violation bugs from \citet{ma_et_al_asplos_2022} for which 
  the \monitorname{} reports an error on the buggy waveform,
  and successfully infers a transaction trace on the fixed waveform. 
  LoC and no. of \code|prot| definitions are for the corresponding \lang{} protocols.
}
\label{fig:bnw_bug_table}
\centering
\footnotesize
\setlength{\tabcolsep}{3pt}
\begin{tabular}{L{4.20cm}rr}
\toprule
\textbf{Bug ID from \citet{ma_et_al_asplos_2022}} & \textbf{LoC} & \textbf{No. of \code|prot|} \\
\midrule
\code|axi-lite-s1| & 85 & 5 \\
\code|axi-stream-s2| & 23 & 2 \\
\code|axis-adapter-s3| & 145 & 3 \\
\code|axi-burst-s4| & 31 & 2 \\
\code|axis-async-fifo-c4| & 32 & 4 \\
\bottomrule
\end{tabular}
\end{table}

\subsubsection{AXI-Stream Protocol Violation Bugs}
\label{sec:bnw_bug_case_studies}

We applied \lang{} to the hardware designs from an existing dataset of bugs in open-source FPGA projects from \citet{ma_et_al_asplos_2022}.
We focus on five \emph{protocol violation bugs} in this dataset:
bugs triggered due to discrepancies between the implementation and the 
protocol's prose specification.
Most protocol violation bugs in this dataset 
occur in AXI-Stream~\cite{arm_axi_stream_2021}, an on-chip communication protocol 
for streaming data.

To use our \monitorname{}, we first wrote
\lang{} specifications expressing the I/O behavior of the DUT 
by consulting the AXI-Stream specification~\cite{arm_axi_stream_2021}.
Although many of these DUTs communicate using standardized protocols, we made 
adjustments to our protocols for each DUT as their I/O behavior 
deviates in situations that are not fully specified by the specification. 
For example, the AXI-Stream specification does not stipulate whether 
the optional \texttt{tkeep} signal (bitmask indicating which bits of 
the data payload are meaningful) should only contain one contiguous
subsequence of 1s, but one DUT implicitly assumes this fact \cite{bnw_s3_bug}.

Each bug in this dataset has a testbench and two versions of the same Verilog module:
a buggy version and a fixed version.
For each bug, we used the testbenches to obtain
two waveforms corresponding to the buggy and correct implementations.
For each of the five protocol violation bugs, the \monitorname{} successfully infers a transaction trace
on the fixed waveform and reports a protocol violation error on the buggy waveform.
Table~\ref{fig:bnw_bug_table} lists details about the \lang{} programs and the \monitorname{}'s behavior on 
the waveforms.

As an example, we focus on one bug from \citet{ma_et_al_asplos_2022} (\texttt{s2} in their artifact).
The DUT is a Xilinx AXI-Stream manager core \cite{zipcpu_xlnxstream2018_3}
which failed to ensure that the \texttt{tlast} signal (indicating the end of a packet) remained stable 
when \texttt{valid = 1} and \texttt{ready = 0} (i.e., during periods of backpressure). 
On the buggy waveform, the \monitorname{} produces an error message indicating 
\texttt{tlast} contains an erroneous value (Figure~\ref{fig:reconstructor_error_example}). On the 
fixed DUT's waveform, the \monitorname{} successfully infers a transaction trace (Figure~\ref{fig:reconstructor_trace_example}).

\begin{figure}
  \begin{subfigure}[b]{\linewidth}
{\catcode`\^=12
\begin{lstlisting}[aboveskip=0pt,belowskip=0pt,basicstyle=\ttfamily\fontsize{8.5}{10.2}\selectfont,moredelim={[is][\color{pasoRed}]{@}{@}}]
@error@: [recv] transaction failed at 1037.5 ns!
   assert_eq(DUT.tlast, is_last);
             ^^^^^^^^^^
   @expected DUT.tlast = 0 but got DUT.tlast = 1@
\end{lstlisting}
}
    \subcaption{
      Error message from the \monitorname{} indicates 
      that the \texttt{tlast} signal is not held stable at 0.
    }
    \label{fig:reconstructor_error_example}
  \end{subfigure}

  \begin{subfigure}[b]{\linewidth}
\begin{lstlisting}[language=paso,aboveskip=0pt,belowskip=0pt,basicstyle=\ttfamily\fontsize{8.5}{10.2}\selectfont]
reset();                  // [time: 0ns -> 12.5ns]
recv(data=1, is_last=0);  // [time: 12.5ns -> 887.5ns]
... // Other transactions omitted
recv(data=6, is_last=0);  // [time: 987.5ns -> 1012.5ns]
\end{lstlisting}
    \subcaption{
      Reconstructor infers transactions from the fixed waveform.
    }
    \label{fig:reconstructor_trace_example}
  \end{subfigure}
  \caption{
    Error messages and inferred transactions produced by the \monitorname{} on the 
    waveforms produced by buggy and fixed versions of an AXI-Stream manager from 
    Ma et al.'s dataset \cite{ma_et_al_asplos_2022}.
  }
  \label{fig:monitor_error_msg}
\end{figure}

\section{Related Work}
\label{sec:related_work}

\paragraph{DSLs for specifying hardware protocols}
Efeu \cite{efeu_eurosys_2025} allows users to write unified 
specifications for the I2C protocol and automatically synthesizes verified device drivers.
Sockeye \cite{sockeye_sosp_2026} transcribes prose specifications from hardware 
manuals into executable specifications, allowing one to prove security properties. 
These efforts are orthogonal to \lang{}: we target on-chip communication protocols
and focus on avoiding driver-monitor discrepancies in testbenches.

\paragraph{Inferring transactions from signals}
Existing hardware testing frameworks (e.g., UVM \cite{uvm_manual}, Cocotb
\cite{cocotb}) allow users to write monitors that recognize transactions 
by observing DUT signal activity. WAL (Waveform Analysis Language)
\cite{wal_aspdac2022, wal_aspdac_2024} enables programmable analysis of waveform traces.
These frameworks require users to manually implement monitors
and drivers for individual protocols, whereas \lang{} 
allows for one centralized protocol specification
that can both drive and monitor RTL designs.

\paragraph{Bugs in RTL designs}
\label{sec:rtl_bugs}
Ma et al. \cite{ma_et_al_asplos_2022} taxonomize common bug patterns in open-source FPGA
designs, while Zonta-Roudes et al. discuss security bugs in implementations of the 
AXI and AXI-Lite protocols \cite{expect_iccad_2024, xray_date_2025}.
Most existing work focuses on functional logic and security bugs in RTL designs, 
whereas \lang{} specifically targets protocol violation bugs.

\section{Conclusion}
\label{sec:conclusion}
Writing separate drivers and monitors to test RTL designs risks inconsistencies.
In \lang{}, the same protocol specification lets us both drive designs and
infer transaction traces from waveforms,
serving as a single source of truth for the RTL design's I/O behavior.


\appendix

\section{Wishbone Specification in \lang{}}
\label{sec:wishbone-appendix}

\begin{lstlisting}[language=paso, breaklines=true, basicstyle=\ttfamily\fontsize{8}{10.2}\selectfont]
// Wishbone protocol specification based on
// the WISHBONE, Revision B.4 Specification.
// We model read/write and burst transactions.
// We do not model the pipelined version.

// Wishbone pin declaration from the server perspective.
interface Wishbone {
    in RST: u1,
    out ACK: u1,
    in ADR: u32,
    in CYC: u1,
    out DAT_I: u32,
    in DAT_O: u32,
    out ERR: u1,
    out RTY: u1,
    in SEL: u4,
    in STB: u1,
    in WE: u1,
    in BTE: u2,
    in CTI: u3,
}

prot reset<DUT: Wishbone>() {
    DUT.RST := 1'b1;
    DUT.CYC := 1'b0;
    DUT.STB := 1'b0;
    step();
}

#[idle]
prot idle_no_cycle<DUT: Wishbone>() {
    DUT.RST := 1'b0;
    DUT.CYC := 1'b0;
    step();
}

#[idle]
prot idle_continue_cycle<DUT: Wishbone>() {
    DUT.RST := 1'b0;
    DUT.CYC := 1'b1;
    DUT.STB := 1'b0;
    step();
}

prot write<DUT: Wishbone>(mask: u4, addr: u32, data: u32) {
    // no RST
    DUT.RST := 1'b0;

    // classic cycle
    DUT.CTI := 3'b0;
    DUT.BTE := X;

    DUT.CYC := 1'b1;
    DUT.STB := 1'b1;

    // we are writing
    DUT.WE := 1'b1;
    DUT.ADR := addr;
    DUT.SEL := mask;
    DUT.DAT_O := data;

    while DUT.ACK == 1'b0 {
        step();
    }

    // acknowledge is now true
    step();
}

prot read<DUT: Wishbone>(mask: u4, addr: u32, data: u32) {
    // no RST
    DUT.RST := 1'b0;

    // classic cycle
    DUT.CTI := 3'b0;
    DUT.BTE := X;

    DUT.CYC := 1'b1;
    DUT.STB := 1'b1;

    // we are reading
    DUT.WE := 1'b0;

    // apply parameters
    DUT.ADR := addr;
    DUT.SEL := mask;
    DUT.DAT_O := X;

    while DUT.ACK == 1'b0 {
        step();
    }

    // check data once ACK is asserted
    assert_eq(DUT.DAT_I, data);
    step();
}

// A burst with only a single item uses a special cti of 7.
// Thus there is no distinction between different address modes.
prot write_burst_single<DUT: Wishbone>(mask: u4, addr: u32,
                                       data: u32) {
    // no RST
    DUT.RST := 1'b0;
    // final burst item
    DUT.CTI := 3'b111;
    DUT.BTE := X;

    DUT.CYC := 1'b1;
    DUT.STB := 1'b1;

    // we are writing
    DUT.WE := 1'b1;

    // apply parameters
    DUT.ADR := addr;
    DUT.SEL := mask;
    DUT.DAT_O := data;

    // wait for ack
    while DUT.ACK == 1'b0 {
        step();
    }

    // acknowledge is now true for one cycle
    step();
}

// A burst with only a single item uses a special cti of 7.
// Thus there is no distinction between the different address modes.
prot read_burst_single<DUT: Wishbone>(mask: u4, addr: u32, data: u32) {
    // no RST
    DUT.RST := 1'b0;

    // final burst item
    DUT.CTI := 3'b111;
    DUT.BTE := X;

    DUT.CYC := 1'b1;
    DUT.STB := 1'b1;

    // we are reading
    DUT.WE := 1'b0;

    // apply parameters
    DUT.ADR := addr;
    DUT.SEL := mask;


    // wait for ack
    while DUT.ACK == 1'b0 {
        step();
    }

    // acknowledge is now true for one cycle and we get the data
    assert_eq(DUT.DAT_I, data);
    step();
}

// A constant address write burst needs at least two data items to be sent.
// Otherwise, this protocol would alias with `write_burst_single`.
// The `[u32]++` type indicates a sequence of u32 elements with a minimum length of two.
prot write_burst_const<DUT: Wishbone>(mask: u4, addr: u32, data: [u32]++) {
    // no RST
    DUT.RST := 1'b0;

    // constant burst cycle (in which case bte is don't care)
    DUT.CTI := 3'b001;
    DUT.BTE := X;

    DUT.CYC := 1'b1;
    DUT.STB := 1'b1;

    // we are writing
    DUT.WE := 1'b1;

    // apply parameters
    DUT.ADR := addr;
    DUT.SEL := mask;

    for d in data {
        DUT.DAT_O := d;

        // indicate end of cycle for last item
        if is_last() {
            DUT.CTI := 3'b111;
        }

        // wait for ack
        while DUT.ACK == 1'b0 {
            step();
        }

        // acknowledge is now true for one cycle
        step();
    }
}

prot read_burst_const<DUT: Wishbone>(mask: u4, addr: u32, data: [u32]++) {
    // no RST
    DUT.RST := 1'b0;

    // constant burst cycle (in which case bte is don't care)
    DUT.CTI := 3'b001;
    DUT.BTE := X;

    DUT.CYC := 1'b1;
    DUT.STB := 1'b1;

    // we are reading
    DUT.WE := 1'b0;

    // apply parameters
    DUT.ADR := addr;
    DUT.SEL := mask;

    for d in data {
        // indicate end of cycle for last item
        if is_last() {
            DUT.CTI := 3'b111;
        }

        // wait for ack
        while DUT.ACK == 1'b0 {
            step();
        }

        // acknowledge is now true for one cycle and we get the data
        assert_eq(DUT.DAT_I, d); 
        step();
    }
}

prot write_burst_increment_linear<DUT: Wishbone>(mask: u4, start_addr: u32, data: [u32]++) {
    // no RST
    DUT.RST := 1'b0;

    // incremental linear burst
    DUT.CTI := 3'b010;
    DUT.BTE := 2'b00;

    DUT.CYC := 1'b1;
    DUT.STB := 1'b1;

    // we are writing
    DUT.WE := 1'b1;

    // apply parameters
    DUT.SEL := mask;

    for d in data {
        DUT.ADR := start_addr + (iter_count::<u30>() ## 2'b00);
        DUT.DAT_O := d;

        // indicate end of cycle for last item
        if is_last() {
            DUT.CTI := 3'b111;
        }

        // wait for ack
        while DUT.ACK == 1'b0 {
            step();
        }

        // acknowledge is now true for one cycle
        step();
    }
}

prot read_burst_increment_linear<DUT: Wishbone>(mask: u4, start_addr: u32, data: [u32]++) {
    // no RST
    DUT.RST := 1'b0;

    // incremental linear burst
    DUT.CTI := 3'b010;
    DUT.BTE := 2'b00;

    DUT.CYC := 1'b1;
    DUT.STB := 1'b1;

    // we are reading
    DUT.WE := 1'b0;

    // apply parameters
    DUT.SEL := mask;

    for d in data {
        DUT.ADR := start_addr + (iter_count::<u30>() ## 2'b00);

        // indicate end of cycle for last item
        if is_last() {
            DUT.CTI := 3'b111;
        }

        // wait for ack
        while DUT.ACK == 1'b0 {
            step();
        }

        // acknowledge is now true for one cycle and we get the data
        assert_eq(DUT.DAT_I, d);
        step();
    }
}


prot write_burst_increment_wrap_4<DUT: Wishbone>(mask: u4, start_addr: u32, data: [u32]++) {
    // no RST
    DUT.RST := 1'b0;
    // incremental linear burst
    DUT.CTI := 3'b010;
    DUT.BTE := 2'b01; // wrap 4

    DUT.CYC := 1'b1;
    DUT.STB := 1'b1;

    // we are writing
    DUT.WE := 1'b1;

    // apply parameters
    DUT.SEL := mask;

    for d in data {
        // we increment the last 2 bits, modulo the 2-bits that are always zero because of the 32-bit interface
        // the reconstructor is able to deal with this complex expression because it simplifies to
        // `start_addr` in the first iteration of the loop.
        // In later iterations of the loop, all bits of `start_addr` are already known
        // and thus the expression evaluates to a concrete value.
        DUT.ADR := start_addr[31:4] ## (start_addr[3:2] + iter_count::<u2>()) ## start_addr[1:0];
        DUT.DAT_O := d;

        // indicate end of cycle for last item
        if is_last() {
            DUT.CTI := 3'b111;
        }

        // wait for ack
        while DUT.ACK == 1'b0 {
            step();
        }

        // acknowledge is now true for one cycle
        step();
    }
}

prot read_burst_increment_wrap_4<DUT: Wishbone>(mask: u4, start_addr: u32, data: [u32]++) {
    // no RST
    DUT.RST := 1'b0;

    // incremental linear burst
    DUT.CTI := 3'b010;
    DUT.BTE := 2'b01; // wrap 4

    DUT.CYC := 1'b1;
    DUT.STB := 1'b1;

    // we are reading
    DUT.WE := 1'b0;

    // apply parameters
    DUT.SEL := mask;

    for d in data {
        // we increment the last 2 bits, modulo the 2-bits that are always zero because of the 32-bit interface
        DUT.ADR := start_addr[31:4] ## (start_addr[3:2] + iter_count::<u2>()) ## start_addr[1:0];

        // indicate end of cycle for last item
        if is_last() {
            DUT.CTI := 3'b111;
        }

        // wait for ack
        while DUT.ACK == 1'b0 {
            step();
        }

        // acknowledge is now true for one cycle and we get the data
        assert_eq(DUT.DAT_I, d);
        step();
    }
}

prot write_burst_increment_wrap_8<DUT: Wishbone>(mask: u4, start_addr: u32, data: [u32]++) {
    // no RST
    DUT.RST := 1'b0;

    // incremental linear burst
    DUT.CTI := 3'b010;
    DUT.BTE := 2'b10; // wrap 8

    DUT.CYC := 1'b1;
    DUT.STB := 1'b1;

    // we are writing
    DUT.WE := 1'b1;

    // apply parameters
    DUT.SEL := mask;

    for d in data {
        // we increment the last 3 bits, modulo the 2-bits that are always zero because of the 32-bit interface
        DUT.ADR := start_addr[31:5] ## (start_addr[4:2] + iter_count::<u3>()) ## start_addr[1:0];
        DUT.DAT_O := d;

        // indicate end of cycle for last item
        if is_last() {
            DUT.CTI := 3'b111;
        }

        // wait for ack
        while DUT.ACK == 1'b0 {
            step();
        }

        // acknowledge is now true for one cycle
        step();
    }
}

prot read_burst_increment_wrap_8<DUT: Wishbone>(mask: u4, start_addr: u32, data: [u32]++) {
    // no RST
    DUT.RST := 1'b0;

    // incremental linear burst
    DUT.CTI := 3'b010;
    DUT.BTE := 2'b10; // wrap 8

    DUT.CYC := 1'b1;
    DUT.STB := 1'b1;

    // we are reading
    DUT.WE := 1'b0;

    // apply parameters
    DUT.SEL := mask;

    for d in data {
        // we increment the last 3 bits, modulo the 2-bits that are always zero because of the 32-bit interface
        DUT.ADR := start_addr[31:5] ## (start_addr[4:2] + iter_count::<u3>()) ## start_addr[1:0];

        // indicate end of cycle for last item
        if is_last() {
            DUT.CTI := 3'b111;
        }

        // wait for ack
        while DUT.ACK == 1'b0 {
            step();
        }

        // acknowledge is now true for one cycle and we get the data
        assert_eq(DUT.DAT_I, d);
        step();
    }
}

prot write_burst_increment_wrap_16<DUT: Wishbone>(mask: u4, start_addr: u32, data: [u32]++) {
        // no RST
    DUT.RST := 1'b0;

    // incremental linear burst
    DUT.CTI := 3'b010;
    DUT.BTE := 2'b11; // wrap 16

    DUT.CYC := 1'b1;
    DUT.STB := 1'b1;

    // we are writing
    DUT.WE := 1'b1;

    // apply parameters
    DUT.SEL := mask;

    for d in data {
        // we increment the last 4 bits, modulo the 2-bits that are always zero because of the 32-bit interface
        DUT.ADR := start_addr[31:6] ## (start_addr[5:2] + iter_count::<u4>()) ## start_addr[1:0];

        // indicate end of cycle for last item
        if is_last() {
            DUT.CTI := 3'b111;
        }

        // wait for ack
        while DUT.ACK == 1'b0 {
            step();
        }

        // acknowledge is now true for one cycle
        step();
    }
}

prot read_burst_increment_wrap_16<DUT: Wishbone>(mask: u4, start_addr: u32, data: [u32]++) {
    // no RST
    DUT.RST := 1'b0;

    // incremental linear burst
    DUT.CTI := 3'b010;
    DUT.BTE := 2'b11; // wrap 16

    DUT.CYC := 1'b1;
    DUT.STB := 1'b1;

    // we are writing
    DUT.WE := 1'b0;

    // apply parameters
    DUT.SEL := mask;

    for d in data {
        // we increment the last 4 bits, modulo the 2-bits that are always zero because of the 32-bit interface
        DUT.ADR := start_addr[31:6] ## (start_addr[5:2] + iter_count::<u4>()) ## start_addr[1:0];

        // indicate end of cycle for last item
        if is_last() {
            DUT.CTI := 3'b111;
        }

        // wait for ack
        while DUT.ACK == 1'b0 {
            step();
        }

        // acknowledge is now true for one cycle and we get the data
        assert_eq(DUT.DAT_I, d); 
        step();
    }
}
\end{lstlisting}


\begin{thebibliography}{58}

%%% the Web address.

\ifx \showCODEN    \undefined \def \showCODEN     #1{\unskip}     \fi
\ifx \showISBNx    \undefined \def \showISBNx     #1{\unskip}     \fi
\ifx \showISBNxiii \undefined \def \showISBNxiii  #1{\unskip}     \fi
\ifx \showISSN     \undefined \def \showISSN      #1{\unskip}     \fi
\ifx \showLCCN     \undefined \def \showLCCN      #1{\unskip}     \fi
\ifx \shownote     \undefined \def \shownote      #1{#1}          \fi
\ifx \showarticletitle \undefined \def \showarticletitle #1{#1}   \fi
\ifx \showURL      \undefined \def \showURL       {\relax}        \fi
% invisible to TeX
\providecommand\bibfield[2]{#2}
\providecommand\bibinfo[2]{#2}
\providecommand\natexlab[1]{#1}
\providecommand\showeprint[2][]{arXiv:#2}

\bibitem[{Altera Corporation}(2018)]%
        {altera_quartus_manual_2018}
\bibfield{author}{\bibinfo{person}{{Altera Corporation}}.}
  \bibinfo{year}{2018}\natexlab{}.
\newblock \bibinfo{title}{Intel Quartus Prime Standard Edition User Guide:
  Design Recommendations, v18.1}.
\newblock
  \bibinfo{howpublished}{\url{https://docs.altera.com/r/docs/683323/18.1/intel-quartus-prime-standard-edition-user-guide-design-recommendations/avoid-combinational-loops}}.
\newblock


\bibitem[{Antmicro}(2022)]%
        {antmicro_2022_wishbone}
\bibfield{author}{\bibinfo{person}{{Antmicro}}.}
  \bibinfo{year}{2022}\natexlab{}.
\newblock \bibinfo{title}{Wishbone Interconnect Burst Mode Benchmark}.
\newblock
  \bibinfo{howpublished}{\url{https://github.com/antmicro/wishbone-interconnect-burst-mode-benchmark}}.
\newblock


\bibitem[{ARM Limited}(1999)]%
        {ARM:AMBA:1999}
\bibfield{author}{\bibinfo{person}{{ARM Limited}}.}
  \bibinfo{year}{1999}\natexlab{}.
\newblock \bibinfo{booktitle}{\emph{{AMBA Specification (Rev 2.0)}}}.
\newblock ARM Limited.
\newblock
\urldef\tempurl%
\url{https://developer.arm.com/documentation/ihi0011/latest/}
\showURL{%
\tempurl}


\bibitem[{Arm Limited}(2021)]%
        {arm_axi_stream_2021}
\bibfield{author}{\bibinfo{person}{{Arm Limited}}.}
  \bibinfo{year}{2021}\natexlab{}.
\newblock \bibinfo{booktitle}{\emph{{AMBA} {AXI}-Stream Protocol
  Specification}}.
\newblock Arm Limited.
\newblock
\urldef\tempurl%
\url{https://developer.arm.com/documentation/ihi0051/latest/}
\showURL{%
\tempurl}
\newblock
\shownote{Issue B, ID040921}.


\bibitem[Baldoni et~al\mbox{.}(2018)]%
        {baldoni_et_al_2018}
\bibfield{author}{\bibinfo{person}{Roberto Baldoni}, \bibinfo{person}{Emilio
  Coppa}, \bibinfo{person}{Daniele~Cono D'Elia}, \bibinfo{person}{Camil
  Demetrescu}, {and} \bibinfo{person}{Irene Finocchi}.}
  \bibinfo{year}{2018}\natexlab{}.
\newblock \showarticletitle{A Survey of Symbolic Execution Techniques}.
\newblock \bibinfo{journal}{\emph{ACM Comput. Surv.}} \bibinfo{volume}{51},
  \bibinfo{number}{3}, Article \bibinfo{articleno}{50} (\bibinfo{year}{2018}).
\newblock


\bibitem[Berlstein et~al\mbox{.}(2023)]%
        {cider_asplos_2023}
\bibfield{author}{\bibinfo{person}{Griffin Berlstein}, \bibinfo{person}{Rachit
  Nigam}, \bibinfo{person}{Christophe Gyurgyik}, {and} \bibinfo{person}{Adrian
  Sampson}.} \bibinfo{year}{2023}\natexlab{}.
\newblock \showarticletitle{Stepwise Debugging for Hardware Accelerators}. In
  \bibinfo{booktitle}{\emph{Proceedings of the 28th ACM International
  Conference on Architectural Support for Programming Languages and Operating
  Systems (ASPLOS 2023), Volume 2}} (Vancouver, BC, Canada).
  \bibinfo{publisher}{Association for Computing Machinery},
  \bibinfo{address}{New York, NY, USA}, \bibinfo{pages}{778–790}.
\newblock
\showISBNx{9781450399166}
\href{https://doi.org/10.1145/3575693.3575717}{doi:\nolinkurl{10.1145/3575693.3575717}}


\bibitem[{Bespoke Silicon Group}(2018)]%
        {basejump_stl}
\bibfield{author}{\bibinfo{person}{{Bespoke Silicon Group}}.}
  \bibinfo{year}{2018}\natexlab{}.
\newblock \bibinfo{title}{{BaseJump STL}: A {SystemVerilog} Standard Template
  Library for Hardware Design}.
\newblock
  \bibinfo{howpublished}{\url{https://github.com/bespoke-silicon-group/basejump_stl}}.
\newblock


\bibitem[Cadar et~al\mbox{.}(2006)]%
        {exe:ccs06}
\bibfield{author}{\bibinfo{person}{Cristian Cadar}, \bibinfo{person}{Vijay
  Ganesh}, \bibinfo{person}{Peter Pawlowski}, \bibinfo{person}{David Dill},
  {and} \bibinfo{person}{Dawson Engler}.} \bibinfo{year}{2006}\natexlab{}.
\newblock \showarticletitle{EXE: Automatically Generating Inputs of Death}. In
  \bibinfo{booktitle}{\emph{ACM Conference on Computer and Communications
  Security (CCS 2006)}} (Alexandria, VA, USA). \bibinfo{pages}{322--335}.
\newblock


\bibitem[Cadar and Sen(2013)]%
        {cadar_sen_cacm_2013}
\bibfield{author}{\bibinfo{person}{Cristian Cadar} {and}
  \bibinfo{person}{Koushik Sen}.} \bibinfo{year}{2013}\natexlab{}.
\newblock \showarticletitle{Symbolic execution for software testing: three
  decades later}.
\newblock \bibinfo{journal}{\emph{Commun. ACM}} \bibinfo{volume}{56},
  \bibinfo{number}{2} (\bibinfo{date}{Feb.} \bibinfo{year}{2013}),
  \bibinfo{pages}{82–90}.
\newblock
\showISSN{0001-0782}
\href{https://doi.org/10.1145/2408776.2408795}{doi:\nolinkurl{10.1145/2408776.2408795}}


\bibitem[Carloni et~al\mbox{.}(2001)]%
        {carloni_et_al_2001}
\bibfield{author}{\bibinfo{person}{Luca~P Carloni}, \bibinfo{person}{Kenneth~L
  McMillan}, {and} \bibinfo{person}{Alberto~L Sangiovanni-Vincentelli}.}
  \bibinfo{year}{2001}\natexlab{}.
\newblock \showarticletitle{Theory of latency-insensitive design}.
\newblock \bibinfo{journal}{\emph{IEEE Transactions on Computer-Aided Design of
  Integrated Circuits and Systems}} \bibinfo{volume}{20}, \bibinfo{number}{9}
  (\bibinfo{year}{2001}), \bibinfo{pages}{1059--1076}.
\newblock
\href{https://doi.org/10.1109/43.945302}{doi:\nolinkurl{10.1109/43.945302}}


\bibitem[Choi et~al\mbox{.}(2025)]%
        {choi_oopsla2025}
\bibfield{author}{\bibinfo{person}{Joonwon Choi}, \bibinfo{person}{Jaewoo Kim},
  {and} \bibinfo{person}{Jeehoon Kang}.} \bibinfo{year}{2025}\natexlab{}.
\newblock \showarticletitle{{Revamping Verilog Semantics for Foundational
  Verification}}.
\newblock \bibinfo{journal}{\emph{Proceedings of the ACM on Programming
  Languages}}  \bibinfo{volume}{9} (\bibinfo{year}{2025}).
\newblock
Issue OOPSLA2.
\href{https://doi.org/10.1145/3763084}{doi:\nolinkurl{10.1145/3763084}}


\bibitem[Christensen et~al\mbox{.}(2021)]%
        {wire_sorts}
\bibfield{author}{\bibinfo{person}{Michael Christensen},
  \bibinfo{person}{Timothy Sherwood}, \bibinfo{person}{Jonathan Balkind}, {and}
  \bibinfo{person}{Ben Hardekopf}.} \bibinfo{year}{2021}\natexlab{}.
\newblock \showarticletitle{Wire sorts: a language abstraction for safe
  hardware composition}. In \bibinfo{booktitle}{\emph{Proceedings of the 42nd
  ACM SIGPLAN International Conference on Programming Language Design and
  Implementation}} (Virtual, Canada) \emph{(\bibinfo{series}{PLDI 2021})}.
  \bibinfo{publisher}{Association for Computing Machinery},
  \bibinfo{address}{New York, NY, USA}, \bibinfo{pages}{175–189}.
\newblock
\showISBNx{9781450383912}
\href{https://doi.org/10.1145/3453483.3454037}{doi:\nolinkurl{10.1145/3453483.3454037}}


\bibitem[Clarke(1976)]%
        {clarke_1976}
\bibfield{author}{\bibinfo{person}{Lori~A. Clarke}.}
  \bibinfo{year}{1976}\natexlab{}.
\newblock \showarticletitle{A Program Testing System}. In
  \bibinfo{booktitle}{\emph{The ACM Annual Conference}}.
\newblock
\href{https://doi.org/10.1145/800191.805647}{doi:\nolinkurl{10.1145/800191.805647}}


\bibitem[{cocotb contributors}(2021)]%
        {cocotb}
\bibfield{author}{\bibinfo{person}{{cocotb contributors}}.}
  \bibinfo{year}{2021}\natexlab{}.
\newblock \bibinfo{booktitle}{\emph{cocotb: A coroutine based cosimulation
  library for writing VHDL and Verilog testbenches in Python}}.
\newblock
\urldef\tempurl%
\url{https://github.com/cocotb/cocotb}
\showURL{%
\tempurl}
\newblock
\shownote{Python-based chip (RTL) verification framework}.


\bibitem[Cook(1971)]%
        {cook_stoc1971}
\bibfield{author}{\bibinfo{person}{Stephen~A. Cook}.}
  \bibinfo{year}{1971}\natexlab{}.
\newblock \showarticletitle{The complexity of theorem-proving procedures}. In
  \bibinfo{booktitle}{\emph{Proceedings of the Third Annual ACM Symposium on
  Theory of Computing}} (Shaker Heights, Ohio, USA)
  \emph{(\bibinfo{series}{STOC '71})}. \bibinfo{publisher}{Association for
  Computing Machinery}, \bibinfo{address}{New York, NY, USA},
  \bibinfo{pages}{151–158}.
\newblock
\showISBNx{9781450374644}
\href{https://doi.org/10.1145/800157.805047}{doi:\nolinkurl{10.1145/800157.805047}}


\bibitem[{courageheart}(2018)]%
        {amba_apb_sram}
\bibfield{author}{\bibinfo{person}{{courageheart}}.}
  \bibinfo{year}{2018}\natexlab{}.
\newblock \bibinfo{title}{{AMBA\_APB\_SRAM}: An {AMBA} {APB} Subordinate {SRAM}
  Core with a {UVM} Testbench}.
\newblock
  \bibinfo{howpublished}{\url{https://github.com/courageheart/AMBA_APB_SRAM}}.
\newblock


\bibitem[De~Moura and Bj\o{}rner(2011)]%
        {de_moura_bjorner_cacm_2011}
\bibfield{author}{\bibinfo{person}{Leonardo De~Moura} {and}
  \bibinfo{person}{Nikolaj Bj\o{}rner}.} \bibinfo{year}{2011}\natexlab{}.
\newblock \showarticletitle{Satisfiability modulo theories: introduction and
  applications}.
\newblock \bibinfo{journal}{\emph{Commun. ACM}} \bibinfo{volume}{54},
  \bibinfo{number}{9} (\bibinfo{date}{Sept.} \bibinfo{year}{2011}),
  \bibinfo{pages}{69–77}.
\newblock
\showISSN{0001-0782}
\href{https://doi.org/10.1145/1995376.1995394}{doi:\nolinkurl{10.1145/1995376.1995394}}


\bibitem[Dobis et~al\mbox{.}(2023)]%
        {dobis2023verification}
\bibfield{author}{\bibinfo{person}{Amelia Dobis}, \bibinfo{person}{Kevin
  Laeufer}, \bibinfo{person}{Hans~Jakob Damsgaard}, \bibinfo{person}{Tjark
  Petersen}, \bibinfo{person}{Kasper Juul~Hesse Rasmussen},
  \bibinfo{person}{Enrico Tolotto}, \bibinfo{person}{Simon~Thye Andersen},
  \bibinfo{person}{Richard Lin}, {and} \bibinfo{person}{Martin Schoeberl}.}
  \bibinfo{year}{2023}\natexlab{}.
\newblock \showarticletitle{{Verification of Chisel Hardware Designs with
  ChiselVerify}}.
\newblock \bibinfo{journal}{\emph{Microprocessors and Microsystems}}
  \bibinfo{volume}{96} (\bibinfo{year}{2023}), \bibinfo{pages}{104737}.
\newblock


\bibitem[{Enjoy Digital}(2019)]%
        {litex_wishbone}
\bibfield{author}{\bibinfo{person}{{Enjoy Digital}}.}
  \bibinfo{year}{2019}\natexlab{}.
\newblock \bibinfo{title}{Litex SoC Framework}.
\newblock \bibinfo{howpublished}{\url{https://github.com/enjoy-digital/litex}}.
\newblock


\bibitem[Fiedler et~al\mbox{.}(2026)]%
        {sockeye_sosp_2026}
\bibfield{author}{\bibinfo{person}{Ben Fiedler}, \bibinfo{person}{Sedan
  Abdelgawad}, \bibinfo{person}{Teymour Aldridge}, \bibinfo{person}{Viktor
  Fukala}, \bibinfo{person}{Jan Häussermann}, \bibinfo{person}{Gamal Hassan},
  \bibinfo{person}{Lars Leuthold}, \bibinfo{person}{Konstantin Lucny},
  \bibinfo{person}{Max Wierse}, \bibinfo{person}{Samuel Gruetter}, {and}
  \bibinfo{person}{Timothy Roscoe}.} \bibinfo{year}{2026}\natexlab{}.
\newblock \showarticletitle{{Sockeye: Bug-finding and proofs for platform
  configurations and hardware based on reference manuals}}. In
  \bibinfo{booktitle}{\emph{SOSP '26: Proceedings of the 32nd ACM Symposium on
  Operating Systems Principles}} (Prague, Czech Republic).
  \bibinfo{publisher}{ACM Press}, \bibinfo{address}{New York, NY, USA}.
\newblock
\urldef\tempurl%
\url{https://arxiv.org/abs/2510.27485}
\showURL{%
\tempurl}
\newblock
\shownote{To appear}.


\bibitem[Forencich(2015)]%
        {bnw_s3_bug}
\bibfield{author}{\bibinfo{person}{Alex Forencich}.}
  \bibinfo{year}{2015}\natexlab{}.
\newblock \bibinfo{title}{AXI4-Stream Adapter Implementation}.
\newblock
  \bibinfo{howpublished}{\url{https://github.com/alexforencich/verilog-axis/commit/9cca78bc7c082b1bff9dd4168fac4841dd47b03b}}.
\newblock


\bibitem[Forencich(2024)]%
        {forencich_verilog_axis}
\bibfield{author}{\bibinfo{person}{Alex Forencich}.}
  \bibinfo{year}{2024}\natexlab{}.
\newblock \bibinfo{title}{{verilog-axis}: {AXI4-Stream} Components for {FPGA}
  Implementation}.
\newblock
  \bibinfo{howpublished}{\url{https://github.com/alexforencich/verilog-axis}}.
\newblock


\bibitem[Foster(2020)]%
        {siemens_2020_fpga_report}
\bibfield{author}{\bibinfo{person}{Harry Foster}.}
  \bibinfo{year}{2020}\natexlab{}.
\newblock \bibinfo{booktitle}{\emph{2020 {Wilson} {Research} {Group} Functional
  Verification Study: {FPGA} Functional Verification Trend Report}}.
\newblock \bibinfo{type}{{T}echnical {R}eport} 82392-C3.
  \bibinfo{institution}{Siemens Digital Industries Software}.
\newblock
\urldef\tempurl%
\url{https://static.sw.cdn.siemens.com/siemens-disw-assets/public/45BqumOhgROPYNeewFFItL/en-US/Siemens%20SW%20Functional%20verification%20study%20FPGA%20trends%20WP%2082392%20C3.pdf}
\showURL{%
\tempurl}
\newblock
\shownote{White paper}.


\bibitem[Godefroid et~al\mbox{.}(2005)]%
        {dart:pldi05}
\bibfield{author}{\bibinfo{person}{Patrice Godefroid}, \bibinfo{person}{Nils
  Klarlund}, {and} \bibinfo{person}{Koushik Sen}.}
  \bibinfo{year}{2005}\natexlab{}.
\newblock \showarticletitle{DART: directed automated random testing}. In
  \bibinfo{booktitle}{\emph{Programming Language Design and Implementation
  (PLDI)}} (Chicago, IL, USA) \emph{(\bibinfo{series}{PLDI '05})}.
  \bibinfo{publisher}{Association for Computing Machinery},
  \bibinfo{address}{New York, NY, USA}, \bibinfo{pages}{213–223}.
\newblock
\showISBNx{1595930566}
\href{https://doi.org/10.1145/1065010.1065036}{doi:\nolinkurl{10.1145/1065010.1065036}}


\bibitem[{GTKWave contributors}({[n.\,d.]})]%
        {gtkwave}
\bibfield{author}{\bibinfo{person}{{GTKWave contributors}}.}
  \bibinfo{year}{[n.\,d.]}\natexlab{}.
\newblock \bibinfo{booktitle}{\emph{{GTKWave}: A Fully Featured {GTK+} Based
  Wave Viewer}}.
\newblock
\urldef\tempurl%
\url{https://gtkwave.sourceforge.net}
\showURL{%
\tempurl}
\newblock
\shownote{Waveform viewer for Unix, Win32, and Mac OSX supporting LXT, LXT2,
  VZT, FST, GHW, and Verilog VCD/EVCD files}.


\bibitem[Herveille(2010)]%
        {Wishbone:2010}
\bibfield{author}{\bibinfo{person}{Richard Herveille}.}
  \bibinfo{year}{2010}\natexlab{}.
\newblock \bibinfo{booktitle}{\emph{{WISHBONE System-on-Chip (SoC)
  Interconnection Architecture for Portable IP Cores}}}.
\newblock OpenCores.
\newblock
\urldef\tempurl%
\url{https://cdn.opencores.org/downloads/wbspec_b4.pdf}
\showURL{%
\tempurl}
\newblock
\shownote{Revision B.4}.


\bibitem[Hsing(2012)]%
        {tiny_aes}
\bibfield{author}{\bibinfo{person}{Homer Hsing}.}
  \bibinfo{year}{2012}\natexlab{}.
\newblock \bibinfo{title}{FreeCores {AES} Core}.
\newblock \bibinfo{howpublished}{\url{https://github.com/freecores/tiny_aes}}.
\newblock
\newblock
\shownote{Pipelined 128-bit AES encryption core}.


\bibitem[IEEE(2020)]%
        {uvm_manual}
\bibfield{author}{\bibinfo{person}{IEEE}.} \bibinfo{year}{2020}\natexlab{}.
\newblock \showarticletitle{IEEE Standard for Universal Verification
  Methodology Language Reference Manual}.
\newblock \bibinfo{journal}{\emph{IEEE Std 1800.2-2020 (Revision of IEEE Std
  1800.2-2017)}} (\bibinfo{year}{2020}), \bibinfo{pages}{1--458}.
\newblock
\href{https://doi.org/10.1109/IEEESTD.2020.9195920}{doi:\nolinkurl{10.1109/IEEESTD.2020.9195920}}


\bibitem[IEEE(2024)]%
        {system_verilog_spec}
\bibfield{author}{\bibinfo{person}{IEEE}.} \bibinfo{year}{2024}\natexlab{}.
\newblock \showarticletitle{IEEE Standard for SystemVerilog--Unified Hardware
  Design, Specification, and Verification Language}.
\newblock \bibinfo{journal}{\emph{IEEE Std 1800-2023 (Revision of IEEE Std
  1800-2017)}} (\bibinfo{year}{2024}), \bibinfo{pages}{1--1354}.
\newblock
\href{https://doi.org/10.1109/IEEESTD.2024.10458102}{doi:\nolinkurl{10.1109/IEEESTD.2024.10458102}}


\bibitem[Kindgren(2018)]%
        {serv}
\bibfield{author}{\bibinfo{person}{Olof Kindgren}.}
  \bibinfo{year}{2018}\natexlab{}.
\newblock \bibinfo{title}{{SERV}: The {SErial} {RISC-V} {CPU}}.
\newblock \bibinfo{howpublished}{\url{https://github.com/olofk/serv}}.
\newblock
\newblock
\shownote{Bbit-serial RISC-V CPU core}.


\bibitem[Kindgren(2026)]%
        {wb_streamer}
\bibfield{author}{\bibinfo{person}{Olof Kindgren}.}
  \bibinfo{year}{2026}\natexlab{}.
\newblock \bibinfo{title}{Wishbone Component for Converting Data Streams to
  Wishbone Transactions}.
\newblock \bibinfo{howpublished}{\url{https://github.com/olofk/wb_streamer}}.
\newblock
\newblock
\shownote{Accessed: 2026-04-21}.


\bibitem[King(1976)]%
        {king_1976}
\bibfield{author}{\bibinfo{person}{James~C. King}.}
  \bibinfo{year}{1976}\natexlab{}.
\newblock \showarticletitle{Symbolic execution and program testing}.
\newblock \bibinfo{journal}{\emph{Commun. ACM}} \bibinfo{volume}{19},
  \bibinfo{number}{7} (\bibinfo{date}{July} \bibinfo{year}{1976}),
  \bibinfo{pages}{385–394}.
\newblock
\showISSN{0001-0782}
\href{https://doi.org/10.1145/360248.360252}{doi:\nolinkurl{10.1145/360248.360252}}


\bibitem[Klemmer and Gro\ss{}e(2022)]%
        {wal_aspdac2022}
\bibfield{author}{\bibinfo{person}{Lucas Klemmer} {and} \bibinfo{person}{Daniel
  Gro\ss{}e}.} \bibinfo{year}{2022}\natexlab{}.
\newblock \showarticletitle{WAL: A Novel Waveform Analysis Language for
  Advanced Design Understanding and Debugging}. In
  \bibinfo{booktitle}{\emph{Proceedings of the 27th Asia and South Pacific
  Design Automation Conference}} (Taipei, Taiwan)
  \emph{(\bibinfo{series}{ASPDAC '22})}. \bibinfo{publisher}{IEEE Press},
  \bibinfo{pages}{358–364}.
\newblock
\showISBNx{9781665421355}
\href{https://doi.org/10.1109/ASP-DAC52403.2022.9712600}{doi:\nolinkurl{10.1109/ASP-DAC52403.2022.9712600}}


\bibitem[Klemmer and Große(2024)]%
        {wal_aspdac_2024}
\bibfield{author}{\bibinfo{person}{Lucas Klemmer} {and} \bibinfo{person}{Daniel
  Große}.} \bibinfo{year}{2024}\natexlab{}.
\newblock \showarticletitle{Towards a Highly Interactive
  Design-Debug-Verification Cycle}. In \bibinfo{booktitle}{\emph{2024 29th Asia
  and South Pacific Design Automation Conference (ASP-DAC)}}.
  \bibinfo{pages}{692--697}.
\newblock
\href{https://doi.org/10.1109/ASP-DAC58780.2024.10473953}{doi:\nolinkurl{10.1109/ASP-DAC58780.2024.10473953}}


\bibitem[Ko{\l}ucki et~al\mbox{.}(2022a)]%
        {antmicro_wb_burst_benchmark}
\bibfield{author}{\bibinfo{person}{Rafa{\l} Ko{\l}ucki},
  \bibinfo{person}{Michael Gielda}, \bibinfo{person}{Karol Guga{\l}a}, {and}
  \bibinfo{person}{J. Szwedyk}.} \bibinfo{year}{2022}\natexlab{a}.
\newblock \bibinfo{title}{Antmicro Wishbone Interconnect Burst Mode Benchmark}.
\newblock
  \bibinfo{howpublished}{\url{https://github.com/antmicro/wishbone-interconnect-burst-mode-benchmark}}.
\newblock


\bibitem[Ko{\l}ucki et~al\mbox{.}(2022b)]%
        {antmicro_wb_bug}
\bibfield{author}{\bibinfo{person}{Rafa{\l} Ko{\l}ucki},
  \bibinfo{person}{Michael Gielda}, \bibinfo{person}{Karol Guga{\l}a}, {and}
  \bibinfo{person}{J. Szwedyk}.} \bibinfo{year}{2022}\natexlab{b}.
\newblock \bibinfo{title}{Antmicro Wishbone Interconnect Burst Mode Benchmark}.
\newblock
  \bibinfo{howpublished}{\url{https://github.com/antmicro/wishbone-interconnect-burst-mode-benchmark/blob/547fa0e9d7256059354559bbbe5d8b95272aab69/test/tests/wb_master.py\#L51}}.
\newblock


\bibitem[Laeufer(2024)]%
        {laeufer_thesis_2024}
\bibfield{author}{\bibinfo{person}{Kevin Laeufer}.}
  \bibinfo{year}{2024}\natexlab{}.
\newblock \emph{\bibinfo{title}{Automated Testing, Verification and Repair of
  RTL Hardware Designs}}.
\newblock \bibinfo{thesistype}{Ph.\,D. Dissertation}. \bibinfo{school}{EECS
  Department, University of California, Berkeley}.
\newblock
\urldef\tempurl%
\url{http://www2.eecs.berkeley.edu/Pubs/TechRpts/2024/EECS-2024-157.html}
\showURL{%
\tempurl}


\bibitem[Leens(2015)]%
        {exostiv_2015_fpga_report}
\bibfield{author}{\bibinfo{person}{Federic Leens}.}
  \bibinfo{year}{2015}\natexlab{}.
\newblock \bibinfo{title}{`{My} {FPGA} Debug and Verification Flow Should Be
  Improved\ldots'}.
\newblock
  \bibinfo{howpublished}{\url{https://www.exostivlabs.com/fpga-debug-flow-should-be-improved/}}.
\newblock
\newblock
\shownote{Archived at
  \url{https://web.archive.org/web/20160128081709/https://www.exostivlabs.com/fpga-debug-flow-should-be-improved/}}.


\bibitem[Lin and Laeufer(2021)]%
        {chiseltest}
\bibfield{author}{\bibinfo{person}{Richard Lin} {and} \bibinfo{person}{Kevin
  Laeufer}.} \bibinfo{year}{2021}\natexlab{}.
\newblock \bibinfo{booktitle}{\emph{chiseltest: The batteries-included testing
  and formal verification library for Chisel-based RTL designs.}}
\newblock
\urldef\tempurl%
\url{https://github.com/ucb-bar/chiseltest}
\showURL{%
\tempurl}


\bibitem[Lööw(2025)]%
        {loow_oopsla25}
\bibfield{author}{\bibinfo{person}{Andreas Lööw}.}
  \bibinfo{year}{2025}\natexlab{}.
\newblock \showarticletitle{{The Simulation Semantics of Synthesisable
  Verilog}}.
\newblock \bibinfo{journal}{\emph{Proceedings of the ACM on Programming
  Languages}}  \bibinfo{volume}{9} (\bibinfo{year}{2025}).
\newblock
Issue OOPSLA1.
\href{https://doi.org/10.1145/3720484}{doi:\nolinkurl{10.1145/3720484}}


\bibitem[Ma et~al\mbox{.}(2022)]%
        {ma_et_al_asplos_2022}
\bibfield{author}{\bibinfo{person}{Jiacheng Ma}, \bibinfo{person}{Gefei Zuo},
  \bibinfo{person}{Kevin Loughlin}, \bibinfo{person}{Haoyang Zhang},
  \bibinfo{person}{Andrew Quinn}, {and} \bibinfo{person}{Baris Kasikci}.}
  \bibinfo{year}{2022}\natexlab{}.
\newblock \showarticletitle{Debugging in the brave new world of reconfigurable
  hardware}. In \bibinfo{booktitle}{\emph{Proceedings of the 27th ACM
  International Conference on Architectural Support for Programming Languages
  and Operating Systems}} (Lausanne, Switzerland)
  \emph{(\bibinfo{series}{ASPLOS '22})}. \bibinfo{publisher}{Association for
  Computing Machinery}, \bibinfo{address}{New York, NY, USA},
  \bibinfo{pages}{946–962}.
\newblock
\showISBNx{9781450392051}
\href{https://doi.org/10.1145/3503222.3507701}{doi:\nolinkurl{10.1145/3503222.3507701}}


\bibitem[{micro-FPGA}(2018)]%
        {engine_v}
\bibfield{author}{\bibinfo{person}{{micro-FPGA}}.}
  \bibinfo{year}{2018}\natexlab{}.
\newblock \bibinfo{title}{{Engine-V}: A {RISC-V} {SoftCPU}/{SoC} for {FPGA}s}.
\newblock \bibinfo{howpublished}{\url{https://github.com/micro-FPGA/engine-V}}.
\newblock


\bibitem[Mohor and Kindgren(2001)]%
        {ethmac}
\bibfield{author}{\bibinfo{person}{Igor Mohor} {and} \bibinfo{person}{Olof
  Kindgren}.} \bibinfo{year}{2001}\natexlab{}.
\newblock \bibinfo{title}{ethmac: Ethernet {MAC} 10/100 {Mbps}}.
\newblock \bibinfo{howpublished}{\url{https://github.com/freecores/ethmac}}.
\newblock


\bibitem[Nigam et~al\mbox{.}(2023)]%
        {filament}
\bibfield{author}{\bibinfo{person}{Rachit Nigam},
  \bibinfo{person}{Pedro~Henrique Azevedo~de Amorim}, {and}
  \bibinfo{person}{Adrian Sampson}.} \bibinfo{year}{2023}\natexlab{}.
\newblock \showarticletitle{Modular Hardware Design with Timeline Types}. In
  \bibinfo{booktitle}{\emph{Proceedings of the 44th ACM SIGPLAN Conference on
  Programming Language Design and Implementation (PLDI)}},
  Vol.~\bibinfo{volume}{7}. \bibinfo{publisher}{Association for Computing
  Machinery}, \bibinfo{address}{New York, NY, USA}, Article
  \bibinfo{articleno}{120}, \bibinfo{numpages}{25}~pages.
\newblock
\href{https://doi.org/10.1145/3591234}{doi:\nolinkurl{10.1145/3591234}}


\bibitem[Nigam et~al\mbox{.}(2026)]%
        {lilac}
\bibfield{author}{\bibinfo{person}{Rachit Nigam}, \bibinfo{person}{Ethan
  Gabizon}, \bibinfo{person}{Edmund Lam}, \bibinfo{person}{Carolyn Zech},
  \bibinfo{person}{Jonathan Balkind}, {and} \bibinfo{person}{Adrian Sampson}.}
  \bibinfo{year}{2026}\natexlab{}.
\newblock \showarticletitle{{Parameterized Hardware Design with
  Latency-Abstract Interfaces}}. In \bibinfo{booktitle}{\emph{ASPLOS '26: 31st
  ACM International Conference on Architectural Support for Programming
  Languages and Operating Systems}}.
\newblock
\href{https://doi.org/10.1145/3779212.3790199}{doi:\nolinkurl{10.1145/3779212.3790199}}


\bibitem[Peter(2023)]%
        {hyperfine_2023}
\bibfield{author}{\bibinfo{person}{David Peter}.}
  \bibinfo{year}{2023}\natexlab{}.
\newblock \bibinfo{booktitle}{\emph{{hyperfine}}}.
\newblock
\urldef\tempurl%
\url{https://github.com/sharkdp/hyperfine}
\showURL{%
\tempurl}


\bibitem[Raj(2023)]%
        {raj_apb}
\bibfield{author}{\bibinfo{person}{Mitu Raj}.} \bibinfo{year}{2023}\natexlab{}.
\newblock \bibinfo{title}{{APB} Manager and Subordinate Implementations in
  {SystemVerilog}}.
\newblock \bibinfo{howpublished}{\url{https://github.com/iammituraj/apb}}.
\newblock


\bibitem[Schemmel et~al\mbox{.}(2020)]%
        {schmemmel_et_al_cav_2020}
\bibfield{author}{\bibinfo{person}{Daniel Schemmel}, \bibinfo{person}{Julian
  B\"{u}ning}, \bibinfo{person}{C\'{e}sar Rodr\'{\i}guez},
  \bibinfo{person}{David Laprell}, {and} \bibinfo{person}{Klaus Wehrle}.}
  \bibinfo{year}{2020}\natexlab{}.
\newblock \showarticletitle{Symbolic Partial-Order Execution for Testing
  Multi-Threaded Programs}. In \bibinfo{booktitle}{\emph{Computer Aided
  Verification (CAV) 2020}} (Los Angeles, CA, USA).
  \bibinfo{publisher}{Springer-Verlag}, \bibinfo{address}{Berlin, Heidelberg},
  \bibinfo{pages}{376–400}.
\newblock
\showISBNx{978-3-030-53287-1}
\href{https://doi.org/10.1007/978-3-030-53288-8_18}{doi:\nolinkurl{10.1007/978-3-030-53288-8_18}}


\bibitem[Schwyn et~al\mbox{.}(2025)]%
        {efeu_eurosys_2025}
\bibfield{author}{\bibinfo{person}{Daniel Schwyn}, \bibinfo{person}{Zikai Liu},
  {and} \bibinfo{person}{Timothy Roscoe}.} \bibinfo{year}{2025}\natexlab{}.
\newblock \showarticletitle{Efeu: generating efficient, verified, hybrid
  hardware/software drivers for I2C devices}. In
  \bibinfo{booktitle}{\emph{Proceedings of the Twentieth European Conference on
  Computer Systems}} (Rotterdam, Netherlands) \emph{(\bibinfo{series}{EuroSys
  '25})}. \bibinfo{publisher}{Association for Computing Machinery},
  \bibinfo{address}{New York, NY, USA}, \bibinfo{pages}{76–93}.
\newblock
\showISBNx{9798400711961}
\href{https://doi.org/10.1145/3689031.3696093}{doi:\nolinkurl{10.1145/3689031.3696093}}


\bibitem[Skarman et~al\mbox{.}(2025)]%
        {surfer_cav_2025}
\bibfield{author}{\bibinfo{person}{Frans Skarman}, \bibinfo{person}{Lucas
  Klemmer}, \bibinfo{person}{Daniel Gro\ss{}e}, \bibinfo{person}{Oscar
  Gustafsson}, {and} \bibinfo{person}{Kevin Laeufer}.}
  \bibinfo{year}{2025}\natexlab{}.
\newblock \showarticletitle{Surfer — An Extensible Waveform Viewer}. In
  \bibinfo{booktitle}{\emph{Computer Aided Verification: 37th International
  Conference, CAV 2025, Zagreb, Croatia, July 23-25, 2025, Proceedings, Part
  IV}} (Zagreb, Croatia). \bibinfo{publisher}{Springer-Verlag},
  \bibinfo{address}{Berlin, Heidelberg}, \bibinfo{pages}{392–404}.
\newblock
\showISBNx{978-3-031-98684-0}
\href{https://doi.org/10.1007/978-3-031-98685-7_19}{doi:\nolinkurl{10.1007/978-3-031-98685-7_19}}


\bibitem[Sutherland(2013)]%
        {sutherland_2013}
\bibfield{author}{\bibinfo{person}{Stuart Sutherland}.}
  \bibinfo{year}{2013}\natexlab{}.
\newblock \showarticletitle{{I'm Still in Love With My X!} (but, do {I} want my
  {X} to be an optimist, a pessimist, or eliminated?)}. In
  \bibinfo{booktitle}{\emph{Design and Verification Conference (DVCon)}}.
  \bibinfo{publisher}{Sutherland HDL, Inc.}, \bibinfo{address}{San Jose, CA}.
\newblock
\urldef\tempurl%
\url{https://dvcon-proceedings.org/wp-content/uploads/im-still-in-love-with-my-x.pdf}
\showURL{%
\tempurl}


\bibitem[Turpin(2003)]%
        {turpin_2003}
\bibfield{author}{\bibinfo{person}{Mike Turpin}.}
  \bibinfo{year}{2003}\natexlab{}.
\newblock \bibinfo{booktitle}{\emph{The Dangers of Living with an {X} (bugs
  hidden in your {V}erilog)}}.
\newblock \bibinfo{type}{{T}echnical {R}eport}. \bibinfo{institution}{ARM
  Ltd.}, \bibinfo{address}{Cambridge, UK}.
\newblock
\newblock
\shownote{Version 1.1, 14 October 2003}.


\bibitem[Wolf and YosysHQ(2015)]%
        {picorv32}
\bibfield{author}{\bibinfo{person}{Claire Wolf} {and}
  \bibinfo{person}{YosysHQ}.} \bibinfo{year}{2015}\natexlab{}.
\newblock \bibinfo{title}{{PicoRV32}: A Size-Optimized {RISC-V} {CPU}}.
\newblock \bibinfo{howpublished}{\url{https://github.com/YosysHQ/picorv32}}.
\newblock
\newblock
\shownote{Open-source RISC-V RV32IMC CPU core licensed under ISC license}.


\bibitem[Yu et~al\mbox{.}(2026)]%
        {yu2026anvil}
\bibfield{author}{\bibinfo{person}{Jason~Zhijingcheng Yu},
  \bibinfo{person}{Aditya~Ranjan Jha}, \bibinfo{person}{Umang Mathur},
  \bibinfo{person}{Trevor~E. Carlson}, {and} \bibinfo{person}{Prateek Saxena}.}
  \bibinfo{year}{2026}\natexlab{}.
\newblock \showarticletitle{{Anvil: A General-Purpose Timing-Safe Hardware
  Description Language}}. In \bibinfo{booktitle}{\emph{ASPLOS '26: 31st ACM
  International Conference on Architectural Support for Programming Languages
  and Operating Systems}}.
\newblock
\href{https://doi.org/10.1145/3779212.3790125}{doi:\nolinkurl{10.1145/3779212.3790125}}


\bibitem[Zhang(2022)]%
        {zhang2022_thesis}
\bibfield{author}{\bibinfo{person}{Keyi Zhang}.}
  \bibinfo{year}{2022}\natexlab{}.
\newblock \emph{\bibinfo{title}{Source-Level Debugging for Hardware Generator
  Frameworks}}.
\newblock \bibinfo{thesistype}{Ph.\,D. Dissertation}. \bibinfo{school}{Stanford
  University}.
\newblock


\bibitem[{ZipCPU (Gisselquist Technology, LLC)}(2018)]%
        {zipcpu_xlnxstream2018_3}
\bibfield{author}{\bibinfo{person}{{ZipCPU (Gisselquist Technology, LLC)}}.}
  \bibinfo{year}{2018}\natexlab{}.
\newblock \bibinfo{title}{{AXI-Stream} master core generated by the {Xilinx
  Vivado} 2018.3 {IP} packager}.
\newblock
  \bibinfo{howpublished}{\url{https://github.com/ZipCPU/wb2axip/blob/master/bench/formal/xlnxstream_2018_3.v}}.
\newblock


\bibitem[Zonta-Roudes et~al\mbox{.}(2025a)]%
        {xray_date_2025}
\bibfield{author}{\bibinfo{person}{Melisande Zonta-Roudes},
  \bibinfo{person}{Nora Hinderling}, {and} \bibinfo{person}{Shweta Shinde}.}
  \bibinfo{year}{2025}\natexlab{a}.
\newblock \showarticletitle{{Xray: Detecting and Exploiting Vulnerabilities in
  Arm AXI Interconnects}}. In \bibinfo{booktitle}{\emph{Design, Automation \&
  Test in Europe Conference (DATE)}}. \bibinfo{pages}{1--7}.
\newblock
\href{https://doi.org/10.23919/DATE64628.2025.10992968}{doi:\nolinkurl{10.23919/DATE64628.2025.10992968}}


\bibitem[Zonta-Roudes et~al\mbox{.}(2025b)]%
        {expect_iccad_2024}
\bibfield{author}{\bibinfo{person}{Melisande Zonta-Roudes},
  \bibinfo{person}{Andres Meza}, \bibinfo{person}{Nora Hinderling},
  \bibinfo{person}{Lucas Deutschmann}, \bibinfo{person}{Francesco Restuccia},
  \bibinfo{person}{Ryan Kastner}, {and} \bibinfo{person}{Shweta Shinde}.}
  \bibinfo{year}{2025}\natexlab{b}.
\newblock \showarticletitle{{eXpect: On the Security Implications of Violations
  in AXI Implementations}}. In \bibinfo{booktitle}{\emph{Proceedings of the
  43rd IEEE/ACM International Conference on Computer-Aided Design (ICCAD
  2024)}}. \bibinfo{publisher}{Association for Computing Machinery},
  \bibinfo{address}{New York, NY, USA}, Article \bibinfo{articleno}{183},
  \bibinfo{numpages}{9}~pages.
\newblock
\showISBNx{9798400710773}
\urldef\tempurl%
\url{https://doi.org/10.1145/3676536.3676844}
\showURL{%
\tempurl}


\end{thebibliography}
\end{document}